\documentstyle[12pt]{article}
\if@twoside\oddsidemargin25mm
\else\oddsidemargin25mm\evensidemargin25mm\marginparwidth25mm\fi%
\begin{document}
\newcommand{\dr}{\raise.3ex\hbox{$\stackrel{\leftarrow}{\partial }$}{}}
\newcommand{\dl}{\raise.3ex\hbox{$\stackrel{\rightarrow}{\partial}$}{}}
\newcommand{\eqn}[1]{(\ref{#1})}
\newcommand{\pl}{(\partial X^\mu )}
\newcommand{\plb}{(\partial X^\nu )}
\newcommand{\plg}{(\partial X^\rho )}
\newcommand{\plr}{(\partial X^\sigma )}
\newcommand{\pls}{(\partial X^\tau )}
\newcommand{\bpl}{(\bar{\partial }X^\mu )}
\newcommand{\ds}{d_{\mu \nu \rho }}
\newcommand{\na}{\nabla }
\newcommand{\half}{{\textstyle\frac{1}{2}}}
\newcommand{\ft}[2]{{\textstyle\frac{#1}{#2}}}
\newcommand{\dkt}{\delta _{KT}}
\newcommand{\QED}{{\hspace*{\fill}\rule{2mm}{2mm}\linebreak}}
\newtheorem{lemma}{Lemma}
\renewcommand{\thelemma}{\thesection.\arabic{lemma}}
\newtheorem{theorem}{Theorem}
\renewcommand{\thetheorem}{\thesection.\arabic{theorem}}
\newtheorem{defn}{Definition}
\renewcommand{\thedefn}{\thesection.\arabic{defn}}
\renewcommand{\theequation}{\thesection.\arabic{equation}}
\newsavebox{\uuunit}
\sbox{\uuunit}
    {\setlength{\unitlength}{0.825em}
     \begin{picture}(0.6,0.7)
        \thinlines
        \put(0,0){\line(1,0){0.5}}
        \put(0.15,0){\line(0,1){0.7}}
        \put(0.35,0){\line(0,1){0.8}}
       \multiput(0.3,0.8)(-0.04,-0.02){12}{\rule{0.5pt}{0.5pt}}
     \end {picture}}
\newcommand {\unity}{\mathord{\!\usebox{\uuunit}}}
\begin{titlepage}
\begin{flushright} KUL-TF-93/23\\ hepth@xxx/9306147  \\
                   June 1993\\
\end{flushright}

\vfill
\begin{center}
{\large\bf Simplifications in Lagrangian BV
quantization\\[7mm] exemplified by the anomalies of chiral
$W_3$ gravity}\\ \vskip 27.mm
{\bf S. Vandoren $^1$ and A. Van Proeyen $^{2,3}$
}\\ \vskip 1cm
Instituut voor Theoretische Fysica
        \\Katholieke Universiteit Leuven
        \\Celestijnenlaan 200D
        \\B--3001 Leuven, Belgium\\[0.3cm]
\end{center}
\vfill
\begin{center}
{\bf Abstract}
\end{center}
\begin{quote}
\small
The Batalin--Vilkovisky (BV) formalism is a useful framework to study
gauge theories. We summarize a simple procedure to find a
a gauge--fixed action in this language and a way to obtain one--loop
anomalies. Calculations
involving the antifields can be greatly simplified by using a theorem on
the antibracket cohomology. The latter is based on properties of a
`Koszul--Tate differential', namely its acyclicity and nilpotency. We
present a new proof for this
acyclicity, respecting locality and covariance of the theory. This theorem
then implies that consistent higher ghost terms in various expressions
exist, and it avoids tedious calculations.
\par
This is illustrated in chiral $W_3$ gravity.
We compute the one--loop anomaly without terms of negative ghost number.
Then the mentioned theorem and the consistency condition imply that the
full anomaly is determined up to local counterterms. Finally we show how to
implement background charges into the BV language in order to
cancel the anomaly with the appropriate counterterms. Again we use the
theorem to simplify the calculations, which agree with previous results.
\vspace{2mm} \vfill \hrule width 3.cm
{\footnotesize
\noindent $^1$ E--mail : Stefan\%tf\%fys@cc3.kuleuven.ac.be \\
\noindent $^2$ E--mail : FGBDA19@cc1.kuleuven.ac.be \\
\noindent $^3$ Onderzoeksleider, NFWO, Belgium}
\normalsize
\end{quote}
\end{titlepage}

\section{Introduction}\label{ss:intro}
The Batalin--Vilkovisky (BV) \cite{BV} method uses the nice
mathematical structure of Poisson--like brackets (antibrackets), canonical
transformations, ... which are the attractive features of the
Hamiltonian language. But one can use the
Lagrangian formalism and in this way keep the advantages of a
covariant formalism. Antifields, $\Phi^*_A$, are the canonically conjugate
variables of the ordinary fields $\Phi ^A$ (which include already the
ghosts). It is well--known that the BV method
is useful for constructing a gauge--fixed action for a broad range of
gauge theories.  It can be applied for all gauge theories which are
known today.  This includes the cases that the algebra of gauge
transformations closes only modulo field equations (`open algebras'),
that there are structure functions rather than constants (`soft
algebras'), that the transformations may not be independent
(`reducible algebras'),~...~. In this language gauge fixing is just a
canonical transformation in the space of fields and antifields
\cite{Siegelgf,anombv,disp,bvsb}.  On the other hand, the formalism also
has been proven to be useful to
study one--loop anomalies \cite{anombv,bvsb,bvleuv}.  In these quantum
aspects, it
is close to the formalism introduced by Zinn--Justin where sources of the
BRST--transformation were introduced \cite{ZinnJustin}. These sources
are the antifields of the BV formalism.  Also from a geometrical
point of view the antifields have a natural origin, as was explained in
\cite{Witten}. More recently, the geometry of the BV formalism
was discussed in \cite{geom}.
However, including these antifields seems unattractable
and very laborious. In this article we will show how one can simplify the
calculations, not including the full load of
antifields in intermediate steps, and still obtain the terms depending on
antifields at the end. We will
make use of the ideas of the Koszul--Tate \cite{KoszulTate} differential
introduced
in \cite{Henn} in this subject. An important theorem, which we will prove
below, will be our main tool. This theorem considers the existence of
a local function $F(\Phi ,\Phi ^*)$ satisfying a
consistency condition. It shows that this function is determined by the
terms of
non--negative ghost numbers (which in some basis, the `classical basis', is
$F(\Phi ,0)$). Further
it shows in how far the remaining terms are fixed by the consistency
condition. In the proofs we follow the ideas of
\cite{Henn}. However, we present new proofs which show the
locality and covariance of the expressions.

We will apply our methods to the calculation of the one--loop anomalies in
chiral $W_3$--gravity. These have already been obtained in
\cite{Hullmatter,SevrinQW3,SSvNMiami,Hullghost,Pope}. The procedure which
we follow is based on a
Pauli--Villars (PV) regularization \cite{PV,anomPV,measure}. The latter can
be done at the level of the action (the path integral), and this
guarantees that one obtains a consistent anomaly. (We will also prove this
directly from the final expression for the one--loop anomaly in
appendix~\ref{app:consista}). Performing the anomaly calculations
for the $W_3$--model including the antifields would lead to
unmanageable expressions. But the theorem mentioned before, allows us
to work with simple expressions, and at the end to
find the complete result. This calculation will demonstrate how the
formalism can be used in more complicated situations than those
discussed previously \cite{anomPV,anombv,measure,W2WZ,exanomBV,ARW}.

We will start this paper with a short review of the BV formalism.
Antifields, the antibracket and
the classical extended action are the ingredients to explain gauge fixing
by a canonical transformation. Further also
the one--loop quantum theory is discussed (section~\ref{ss:ingrBV}). We
repeat the essential formulae to obtain anomalies, and explain why
antifields have to be included. Inclusion of background fields becomes
unnecessary. Section~\ref{s:adbkta} is more formal.
In section~\ref{ss:BVdefn}  we define some important technical ingredients
like locality, stationary surface and properness of the extended action.
In section~\ref{ss:afnKT} we introduce the Koszul--Tate differential and in
section~\ref{ss:KTacycl} we prove its
acyclicity, respecting the locality and covariance of the theory.
Then we come to our main theorem, (section~\ref{ss:ABcoho}), where we prove
the equivalence between the antibracket cohomology and the weak BRST
cohomology.
This theorem can be applied for the analysis of anomalies.
Indeed, in section~\ref{ss:W3} we calculate the one--loop anomaly for
chiral $W_3$ gravity using the simplifications. This calculation is a clear
example of the general method. Finally, in the section~\ref{ss:backgrch},
we introduce background charges in order to cancel the anomaly. We show how
this modifies the usual expansion of the quantum master equation, but
it still fits in the general framework. The theorem on the antibracket
cohomology
can again be used to simplify the analysis. In the concluding section we
will also comment on the higher loop anomalies.

\section{The essential ingredients of the BV--formalism}
\setcounter{equation}{0}
\label{ss:ingrBV}
We will give here a summary of the way in which we use the
BV--formalism. Our methods are now simpler and more transparant than the
original ones of Batalin and Vilkovisky. A short review has been given in
\cite{bvsb}, which will be supplemented in \cite{bvberk}. A complete review
will be given in \cite{bvleuv}.
\subsection{The classical formalism}
We denote by $\{\Phi ^A\}$ the complete set of fields.
It will includes the ghosts for all the gauge symmetries, and possibly
auxiliary fields introduced for
gauge fixing. We will see below that sometimes we do not need more than the
classical fields and the ghosts.  Then
one doubles the space of field variables by
introducing antifields $\Phi ^*_A$, which play the role of canonical
conjugate variables with respect to a Poisson--like structure. This is
defined by means of `antibrackets', whose canonical structure is
\begin{equation}
(\Phi ^A,\, \Phi ^*_B) =\delta ^A{}_B \ ;\qquad (\Phi ^A,\Phi ^B)=(\Phi
^*_A,\Phi ^*_B)=0\ . \label{canbr} \end{equation}
The antifields $\Phi ^*_A$ have opposite statistics than their canonically
conjugate field $\Phi ^A$.
In general the antibracket of $F(\Phi ^A,\Phi ^*_A)$ and
$G(\Phi ^A,\Phi ^*_A)$ is defined by
\begin{equation}
(F,G) =  F\dr_A\cdot\dl{}^A G - F\dr{}^A \cdot\dl_A G\ ,
\label{abracket} \end{equation}
using the notations
\begin{equation}
\partial _A=\frac{\partial }{\partial \Phi ^A}\ ;\qquad
\partial ^A=\frac{\partial }{\partial \Phi ^*_A},
\end{equation}
and $\dr$ and $\dl$ stand for right and left derivatives
acting on the object before, resp. behind, the symbol $\partial $. The
separating symbol $\cdot$ is often useful to indicate up to where the
derivatives act, if they are not enclosed in brackets.
Note that this antibracket is a fermionic operation, in the sense that the
statistics of the antibracket $(F,G)$ is opposite to that of $FG$. The
antibracket also satisfies some graded Jacobi--relations:
\begin{equation}
(F,(G,H))+ (-)^{FG+F+G}(G,(F,H))=((F,G),H). \label{Jacobi}
\end{equation}

We assign {\bf ghost numbers} to fields and antifields. These are
integers such that \begin{equation}
gh(\Phi ^*)+gh(\Phi )=-1\ ,\label{ghfaf}
\end{equation}
and therefore the
antibracket \eqn{abracket} raises the ghost number by 1.

We will often perform canonical transformations in this space of fields and
antifields \cite{BVcan}. These are the transformations such that the
new basis again
satisfies \eqn{canbr}. We will also always respect the ghost numbers. It is
clear that interchanging the name field and antifield of a canonical
conjugate pair ($\phi '=\phi ^*$ and $\phi '^*=-\phi $) is such a
transformation. The new
antifield has the ghost number of the old field. From \eqn{ghfaf} we see
then that there is always a basis in which all fields have positive or zero
ghost numbers, and the antifields have negative ghost numbers. We will
often use that basis. It is the natural one from the point of view of the
classical theory, and therefore we will denote it as the `classical
basis'. We will see below that it is not the most convenient
from the point of view of the path integral.

One defines an `extended action', $S(\Phi ^A,\Phi ^*_A)$, of ghost number
zero, whose
antifield independent part $S(\Phi ^A,0)$ is at this point the classical
action, and which satisfies the {\it master equation}
\begin{equation}
(S,S)=0\ .\label{SS0}
\end{equation}
This equation contains the statements of gauge invariances of the classical
action, their algebra, closure, Jacobi identities, ...~.
In the enlarged field space, gauge fixing
is obtained by a canonical transformation (see example below). This
transformation
is chosen such that the new antifield--independent part of $S$
has no gauge invariances. In this new basis some antifields will have
positive or zero ghost numbers, so it is not any more of the type mentioned
above. Choosing different
canonical transformations satisfying the above requirement amounts to
different gauge choices.

A simple example is 2--dimensional chiral gravity. The classical action is
\begin{equation}
S_0=\int d^2x\, \left[ -\half \partial X^\mu \cdot \bar \partial X^\mu
+\half h\,\partial X^\mu \cdot\partial X^\mu \right] \ .\label{S0grav}
\end{equation}
In the extended action appears a ghost $c$ related to the reparametrization
invariance. The fields are then $\Phi ^A=\{ X^\mu ,h,c\}$. The
extended action is \begin{eqnarray}
&&S=S_0+ \int d^2x\left[ X ^*_\mu \, c\partial X^\mu   +
h^*\left(\bar \partial -h\partial +(\partial h)\right) c\right.\nonumber\\
&&\hspace{3cm} \left.-c^*\,c\,\partial c\right] \ .\label{SextW2}
\end{eqnarray}
Added to the classical action, one finds here the antifields
multiplied with the transformation rules of the classical fields in their
BRST form. One may then check with the above
definitions
that the vanishing of $\left. (S,S)\right|_{\Phi ^*=0}$ expresses the gauge
invariance. The second line contains in the same way the BRST
transformation of the ghost. It is clear that this is determined by the
previous line and \eqn{SS0} (BRST invariance). This is a
trivial example of a principle which we want to stress in this article.
Often antifield dependent terms are already determined by requirements as
\eqn{SS0}, such that one does not have to specify them explicitly, and one
knows that a solution exist without having to make tedious checks of the
higher ghost terms\footnote{Some may remember the checks of quintic ghost
terms in the BRST transformations of supergravity actions.}. In order that
these terms are uniquely
defined (up to canonical transformations) we will need a requirement of
`properness', which in simple words means that ghosts are introduced
for {\it all} the gauge invariances, with the appropriate terms in the
extended action. An exact definition will be given in
section~\ref{ss:BVdefn}.

In this example gauge fixing is obtained by the canonical transformation
where $h$ and $h^*$ are replaced by $b$ and $b^*$:
\begin{equation}
b=h^*\ ;\qquad b^*=-h\ .\label{bhcantr}
\end{equation}
One checks then that the part of $S$ depending only on the new `fields',
i.e. $X^\mu $, $b$ and $c$, which is
$\left[- \half \partial X^\mu \cdot \bar \partial X^\mu  + b\bar
\partial c \right] $,
has no gauge invariances. $b$, which has its origin as antifield of the
classical field $h$, is now considered as a field (the antighost) of
negative ghost number. In more complicated
situations one may first need to include `non--minimal' fields, i.e. other
fields than classical or ghost fields, to be able to
perform a canonical transformation such that the action in the new fields
is `gauge--fixed'.  When quantizing, we will work in this basis,
to which we will refer as the `gauge--fixed basis'.
It is such that the extended action takes the form
\begin{equation}
S(\Phi ,\Phi ^*)=S_{gauge-fixed}(\Phi )+\mbox{   antifield--dependent
terms}\ .
\end{equation}
`Gauge fixed' means that in the new definitions of fields the matrix of
second derivatives w.r.t.  fields, $S_{AB}$, is non--singular when setting
the field equations equal to zero. For more details on this procedure we
refer to \cite{BV,disp,bvsb,bvleuv}.

\subsection{The one--loop theory} \label{ss:oneloop}
This gauge--fixed action is still
a classical action in the sense that no loops ($\hbar $ corrections) are
included. The full quantum (extended) action $W(\Phi ,\Phi ^*)$ can be
expanded in powers of $\hbar$:
\begin{equation}
W=S+\hbar M_1 +\hbar ^2M_2 +...\ .  \label{QA}
\end{equation}
In the previous subsection we discussed the form of $S$, but there is no
reason to assume that this would have no $\hbar $ dependent corrections,
which are the $M_i$. For a local field theory we want also integrals of
local functions for these $M_i$. Note that as we are discussing the
extended action, this includes at once the quantum corrections to the
transformation laws. The expansion \eqn{QA} is the
usual one, but we will see in section~\ref{ss:backgrch} that terms of order
$\sqrt{\hbar }$ can appear.

This quantum action appears in the path integral
\begin{equation}
 Z(J, \Phi ^*)=
\int {\cal D} \Phi \exp \left( \frac{i}{\hbar }W(\Phi,\Phi^*)
+J(\Phi)\right) \ ,\label{pathint}
\end{equation}
where we have introduced sources $J(\Phi )$, and in this subsection we use
the gauge--fixed basis. The gauge fixing which we discussed, can be seen
as the procedure to select out of the $2N$ variables, $\Phi ^A$ and $\Phi
^*_A$, the $N$ variables over which one integrates in this integral (the
`Lagrangian submanifold'). To define the path integral
properly one has to discuss regularization, which can be seen as a way to
define the measure. In gauge theories, in general one can
not find a regularization which respects all the gauge symmetries.
Then it is possible that symmetries of the classical theory are not
preserved in the quantum theory. Anomalies are the expression of this
non--invariance. If there are no anomalies, then the quantum theory does
not depend on the gauge fixing, i.e. on the particular choice of variables
$\Phi ^A$ used for the integral in \eqn{pathint} (as long as
$S_{AB}$ is non--singular). That does not hold when there are anomalies.
In that case the quantum theory will have a different content than the
classical theory. One can obtain in this way induced theories,
which from our point of view are theories where antifields become
propagating fields (this point will be explained in \cite{bvberk}).
In \cite{anombv} it was shown how to use the BV framework to investigate
the anomaly structure. For a theory to be free of anomalies, $W$ has to be
a solution of the master equation
\begin{equation}
(W,W)=2i\hbar \Delta W\ ,   \label{ME}
\end{equation}
where
\begin{equation}
\Delta =(-)^A\dl _A\dl ^A\,. \label{BOX}
\end{equation}
In powers of $\hbar$ the first two equations (zero--loop and one--loop)
are \begin{eqnarray}
&&(S,S) = 0\nonumber\\
&&i{\cal A}\equiv i\Delta S-(M_1,S)=0\ . \label{MEC}
\end{eqnarray}
The first one is the (classical) master equation discussed before. The
second one is an equation for $M_1$. In a local field theory we will
moreover demand that $M_1$ is an integral of a local function (exact
definitions are given
in section~\ref{ss:BVdefn}). If there does not exist such an $M_1$ then
${\cal A}$ is called the anomaly. It is clearly not uniquely defined,
as $M_1$ is arbitrary. It satisfies
\begin{equation}
({\cal A},S)=0\ ,\label{WZccS}
\end{equation}
which is a reformulation of the Wess--Zumino
consistency conditions \cite{WZcc}.
But as mentioned, we need a regularization procedure.  In the expressions
above, we notice this because on any integral of a local function, the
$\Delta $ operator is proportional to $\delta (0)$.

To do so within the context of path
integrals, one can not use dimensional regularization as there is no good
definition of the action in arbitrary (non--integer) dimension.
The method of \cite{anomPV,anombv,measure,bvsb,bvleuv} is based on
Pauli--Villars regularization \cite{PV}. This can be done in the path
integral and in this way one obtains consistent anomalies. We will use this
method in section~\ref{ss:W3}, and will now briefly repeat the essential
features.

The new ingredient is a mass matrix which is introduced for the PV fields:
$T_{AB}$, such that
\begin{equation}
S_M^{PV}=-\ft12 M^2 \Phi_{PV} ^A T_{AB}\Phi_{PV} ^B\ ,  \label{SMPV}
\end{equation}
where $\Phi ^A_{PV}$ are the PV partners of all the fields.
This matrix should be invertible, at least in the set of propagating
fields. Its inverse serves as index--raising metric in the space of fields.
The regulator, ${\cal R}$, is defined from the (non--singular) matrix of
second derivatives of the extended action
\begin{equation}
{\cal R}^A{}_{B}=(T^{-1})^{AC}\,S_{CB}\ ;\qquad S_{AB}=\dl_A S \dr_B\
.\label{regulator} \end{equation}
Note that in general $S_{AB}$ contains antifields.
The anomaly is then obtained as the non--invariance of the mass term
\eqn{SMPV} after integrating over the PV fields. This leads to
\begin{equation}
\Delta S=\lim_{M^2\rightarrow \infty}Tr\left[J\,\frac{1}{1-{\cal R}
/M^2}\right] \ , \label{SSMPV} \end{equation}
where $J$ is given in terms of the transformation matrix $K$,
the derivative of the extended action w.r.t. an antifield and a field, as
\begin{equation}
J^A{}_B=K^A{}_B+\ft12 (T^{-1})^{AC}\left(T_{CB},S\right)(-)^B\ ;\qquad
K^A{}_B = S^A{}_B\ .\label{defK}
\end{equation}
Note that in general $S_{AB}$ contains antifields (even for closed
algebras). Also the other
matrices may be antifield dependent, and thus $\Delta S$
also contains antifields. After some further steps, we may write this
regularized value of $\Delta S$ also as
\begin{equation}
\Delta S=\lim_{M^2\rightarrow \infty}Tr\left[J\,\exp({\cal R}/M^2)\right]\
, \label{BOXS} \end{equation}
The regularized value thus depends on the choice of the matrix $T$.
Different $T$'s correspond to different regularization schemes.
In our example we will use matrices $T$ which have neither fields nor
antifields, and thus $J=K$. Sometimes
it is useful to take a field--dependent mass matrix \cite{measure}.
In appendix~\ref{app:consista} we explicitly check that this expression
for $\Delta S$ satisfies the consistency condition
\begin{equation}
(S,\Delta S)=0\ .          \label{consdelS}
\end{equation}
Alternative formulations of the expression for the anomaly are given in
\cite{JGJPII} (see also appendix~\ref{app:consista}). They show that there
is always an $M_1$ such that \eqn{MEC} is satisfied. However,
this expression is in general non--local. Anomalies appear if
no local expression $M_1$ can be found satisfying that
equation.\vspace{1cm}

To evaluate the traces, one can put the operator between a basis of
plane waves. So
\begin{eqnarray}
Tr\left[K\exp{\frac{{\cal R}}{M^2}}\right]&=& \int d^dx\,\int d^dy \delta
(x-y)K(x) e^{{\cal R}_x/M^2} \delta (x-y)\nonumber\\
&=&
\int d^dx\int \frac{d^dk}{(2\pi )^d}e^ {-ikx}K e^{{\cal R}_x/M^2}
e^{ikx}
\end{eqnarray}
Then one pulls the $e^{ikx}$ to the left, replacing derivatives by
$\partial +ikx$, and one takes the trace. However, this whole procedure is
included in the results of the heat kernel method \cite{heatk}. The
heat kernel is the expression
\begin{equation}
e^{t{\cal R}_x}\delta (x-y)= G(x,y;t;\Phi ) \ ,\label{defGdel}
\end{equation}
which we thus need for $t=1/M^2$. This has been considered for general
second order differential operators ${\cal R}$
\begin{equation}
{\cal R}_x[\Phi ]= {1\over {\sqrt g}}\left( \partial_\alpha \unity +{\cal
Y}_\alpha  \right)
{\sqrt g}\, g^\alpha {}^\beta  \left( \partial_\beta \unity +{\cal Y}_\beta
\right) + E  \ , \label{formcR}
\end{equation}
where $\Phi =\{g^{\alpha \beta  },{\cal Y}_\alpha  ,E\}$. The latter two
can be matrices in an internal space. The restriction here is that
the part which contains second derivatives
is proportional to the unit matrix in internal space. (Note that in the
application for $W_3$, which will be considered in section~\ref{ss:W3}, we
will have a more general regulator, but we will be able to reduce it to
this case by some expansion). Further in principle $g$ should be a positive
definite matrix. For Minkowski space, we have to perform first a Wick
rotation, which introduces a factor $-i$ in the final expression, and thus
we have to use
\begin{equation}
Tr\left[K\exp{\frac{{\cal R}}{M^2}}\right]=-i \int d^dx\,\int d^dy \delta
(x-y)K(x) G\left(x,y;\frac{1}{M^2}\right) \ .
\end{equation}

The `early time' expansion of this heat kernel is as follows
\begin{equation}
G(x,y;t;\Phi )= \frac{\sqrt{g(y)}\,\Delta ^{1/2}(x,y)}{ (4\pi t )^{d/2}}
e^{-\sigma (x,y) /2t } \sum_{n=0}a_n(x,y,\Phi ) t^n\ ,\label{Gexpbn}
\end{equation}
where $g=|det\ g_{\alpha \beta }|$, and
 $\sigma (x,y)$ is the `world function', which is discussed at length
in \cite{Synge}. It is half the square of the geodesic distance between $x$
and $y$
\begin{equation}
\sigma (x,y)=\frac{1}{2}g_{\alpha \beta }(y-x)^\alpha
(y-x)^\beta  +{\cal O}(x-y)^3\ .
\end{equation}
Further, $\Delta (x,y)$ is defined from the `Van Vleck--Morette
determinant' (for more information on this determinant, see
\cite{VisservV})
\begin{equation}
{\cal D}(x,y) =\left|det\left( -\frac{\partial ^2\sigma }{\partial x^\alpha
\partial y^\beta  }\right) \right| \end{equation}
as
\begin{equation}
{\cal D}(x,y) =\sqrt{g(x)}\,\sqrt{g(y)}\,\Delta (x,y)\ .
\end{equation}
At coincident points it is 1, and its first derivative is zero.

The `Seeley--DeWitt' coefficients $a_n(x,y,\Phi )$ have been obtained using
various methods for the most important cases. In two dimensions we only
need $a_0$ and $a_1$. For most applications we only need their value
and first and second
derivatives at coincident points. Let us still note that this expansion
also appears in quantum mechanics,  and vice versa
results of quantum mechanics can be used to prove parts of the heat kernel
expansion \cite{FBPvN}. \vspace{1cm}

If the calculations are done in a specific gauge, one can not see at the
end whether
one has obtained gauge--independent results. Therefore one often includes
parameters in the choice of gauge, or background fields. The question
then remains whether this choice was `general enough' to be able to trigger
all possible anomalies. E.g. for the above example the gauge $h=0$ would
not show the anomaly, which is of the form $\int d^2x\, c\,\partial ^3h$.
Instead, a
gauge choice $h=H$, where the latter is a background field, is sufficient.
For Yang--Mills gauge theories on the other hand, it is difficult to see
how much
generalization of the gauge $\partial ^\mu A_\mu =0$ is necessary to obtain
all possible anomalies. In the BV formalism, no background fields are
necessary, but one keeps
the antifield--dependent terms through all calculations. The anomalies are
reflected at
the end by dependence on these antifields. In our example, we would obtain
an anomaly $\int d^2x\, c\,\partial ^3b^*$. In this formalism it
is then also clear how anomalies change by going to other gauges. The
relation is given by canonical transformations.

\section{Antifield dependence by Koszul--Tate acyclicity}
\label{s:adbkta}
\setcounter{equation}{0}

Inclusion of antifields in all calculations can make them very difficult
and tedious. In this section we will  show that in various expressions, in
particular for the anomaly, the knowledge of the part which does not depend
on fields of negative ghost number, determines the full expression up to
certain counterterms. Indeed, the knowledge that the anomaly satisfies the
consistency condition \eqn{consdelS} gives extra information. We will here
investigate what we can derive from it.
In general the problem which we have
to solve, is to determine a quantity $F(\Phi^A ,\Phi ^*_A)$ restricted by
\begin{equation}
(F,S)=0 \ ,\label{FS0}
\end{equation}
when we know $F^0$, which is the part of $F$ where all fields with
negative ghost number are omitted. In this section we will use the
`classical basis', where all the fields at negative ghost number are
considered to be the antifields, so $F^0=F(\Phi ^A,0)$. We
will see that the problem only has a solution if $F^0$ satisfies a certain
condition. In that case all the other terms of $F$ can be determined
up to terms
$(G,S)$ using a certain expansion which we will define below.  We thus
solve a cohomology problem of the nilpotent operator ${\cal
S}F=(F,S)$, knowing the antifield--independent part of $F$.
In section~\ref{ss:W3} we will illustrate how this simplifies the
calculation of the one--loop anomaly for $W_3$--gravity.

The essential ingredient in the proofs is the
Koszul--Tate (KT) differential, working on the space of antifields.  This
operator and its properties have been studied in detail in
\cite{Henn}. It has been used first of all to prove the
existence and uniqueness of a solution to the master equation.

Most of the statements in this section have been derived already in
\cite{Henn}. The difference in the derivation is that in these articles
it was
assumed that in gauge theories, the field equations can be split in
dependent and independent ones. E.g. in Maxwell theory,
the field equations $y_\mu \equiv \partial ^\nu F_{\mu \nu }=0$ are four
equations, which are dependent: $\partial ^\mu y_\mu =0$. It is indeed
true that these can be split in three independent ones, and one dependent
on the others. However one can not do this splitting in a local (and/or
covariant) way. There
is no local expression of one of the constraints in terms of the others.
If that were possible, then we could just drop the corresponding
field and the symmetry.
In our proofs we will always avoid such splittings. We will only
work with a set of functions, to be denoted by the set of `local functions'
(see subsection~\ref{ss:BVdefn}), which does not allow such a
split. Therefore all the quantities which we construct will be local
functions (and possibly covariant under a rigid symmetry group).
We are then able to prove the statements about the cohomology for
these local functions.

There are of course certain weak conditions on the theories under
considerations. E.g. the field equations should allow at least one
solution. Further conditions were referred to in \cite{BVexist,Henn} as
defining `regular theories'. We will extend the set of theories
for which the theorems are applicable. The previous definition of `regular
theories' excluded already the simple 2--dimensional
theories which we use as examples here. We will introduce the notion of
evanescent functions to be added to the functions on the stationary
surface.

It is clear that we first need a set of definitions
(subsection~\ref{ss:BVdefn}).
We define the `stationary surface', using `classical' field equations,
and specify the meaning of the `local functions' which we
consider. Also the `properness condition' will be discussed.
Then we will be able to define our expansion
and the KT differential in the second subsection. The main property of the
KT differential, its acyclicity,
will be the subject of subsection~\ref{ss:KTacycl}, where we give
the essential ideas, while the technical proofs are in
appendix~\ref{app:lemprKT}. Similarly, the ideas of the antibracket
cohomology are in subsection~\ref{ss:ABcoho}, while the corresponding
proofs are in appendix~\ref{app:ABcoho}.

We will work here in general under the
assumption that there is an extended action $S(\Phi ,\Phi ^*)$, which
satisfies the master equation \eqn{SS0}. However, with slight modifications
the strategy of these proofs can be used also to prove first the
KT--acyclicity before we have the extended action. Then this provides the
proof of the existence and uniqueness of the extended action as in
\cite{Henn}. Our improvements thus apply also to these statements, and the
proofs will be presented in that way in \cite{bvleuv}.

\subsection{The technical ingredients.}
\label{ss:BVdefn}

As mentioned, we use the `classical basis' where antifields have negative
ghost numbers, and fields have zero or positive ghost numbers.
For further use, we now give special names to the fields of each ghost
number. The fields of ghost number zero are denoted by $\phi ^i$. Those of
ghost number 1 are denoted by $c^a$, and those of ghost number $k+1$ are
written as $c^{a_k}$. (In this way the index $i$ can also be denoted as
$a_{-1}$ and for $c^a$ we can also write $c^{a_0}$).
\begin{equation}
\{\Phi ^A\}\equiv \{ \phi ^i,\,c^a,\,c^{a_1},\ldots \} \ .
\end{equation}

When one starts from a classical action, one directly obtains the above
structure, where $\phi ^i$ are the classical fields, $c^a$ are the ghosts,
and the others are `ghosts for ghosts'. In fact, the structure below
with the KT differential is used in the proof that starting from a
classical
action one can always construct solutions of the master equation with zero
ghost number \cite{BVexist,Henn,bvleuv}.

The antifields thus have negative ghost number and we define the
{\bf antifield number} ($afn$) as
\begin{equation}
afn(\Phi ^*_A)=-gh(\Phi ^*_A) >0 \ ;\qquad afn(\Phi ^A)=0\ .
\end{equation}
With the above designation of names to the different fields we thus have
\begin{equation}
afn(\phi ^*_i)=1\ ;\qquad afn(c^*_{a_k})=k+2\ .
\end{equation}
Then every expression can be expanded in terms with definite antifield
number. E.g. for the extended action, we can define
\begin{equation}
S=\sum_{k=0} S^k\ ;\qquad S^k=S^k\left(c^*_{a_{k-2}},c^*_{a_{k-3}},\ldots
,\phi ^i,\ldots ,c^{a_{k-1}}\right)\ ,  \label{Skdep}
\end{equation}
where the range of fields and antifields which can occur in each term
follows from the above definitions of antifield and ghost numbers, and the
requirement $gh(S)=0$.\vspace{1cm}

When constructing a theory, one should
specify the set of functions of $\phi ^i$,
denoted by ${\cal F}^0$, which is considered.
An important aspect of the quantization procedure which we present here is
that we consider {\bf `local
functions'}. This means that they depend on $\phi^i$ and a
finite number of their derivatives at one space--time point (no integrals).
In general we need
more restrictions, which depend on the theory. E.g. one should specify
whether a square root of the field is in the set ${\cal F}^0$. This set
should contain at least the fields themselves, and other functions which
appear in the action and transformation rules. For some applications
one may also consider non--local functions (see e.g. \cite{W2WZ,ARW}).
In each case it will
be important to define exactly what is meant by the set of functions
${\cal F}^0$. For convenience we will call this set `the local
functions'.

$S^0$ depends only on $\phi ^i$. This is `the classical action'. From
this action we obtain the classical field equations
\begin{equation}
y_i\equiv S^0\dr_i \approx 0\ .
\end{equation}
The field configurations which satisfy
these field equations form the {\bf stationary surface}. We define the
symbol $\approx 0 $ (weakly zero) by
\begin{equation}
F\approx 0 \ \Longleftrightarrow\ F=y_i G^i \ , \label{approx}
\end{equation}
where $G^i$ are some local and regular functions $\in {\cal F}^0$.

In some cases one can define ${\cal F}^0$ such that any
function in this set which vanishes on the stationary surface is
proportional to the field equations, and thus  $\approx
0$. Then the theory is called {\it regular}. This is not always the case.
In our example of 2--dimensional chiral gravity,
\eqn{S0grav}, the field equations are \begin{eqnarray}
y_\mu &=&-\bar \partial \partial X^\mu +\partial \left( h(\partial
X^\mu ) \right)\approx 0\nonumber\\
y_h &=& -\ft12 \partial X^\mu \cdot\partial X^\mu \approx 0\ .
\end{eqnarray}
Now $\partial X^\mu $ vanishes on the stationary surface, but it
is not proportional to a field equation. Therefore we say that $\partial
X^\mu $ is not weakly zero. Functions vanishing on the stationary
surface, but not weakly zero, can be called `evanescent functions'.

Within the set of functions ${\cal F}^0$ the relation $\approx $, defines
equivalence classes. These are the functions on the stationary
surface and the classes of evanescent functions.
Further on, we will use the terminology `functions on the
stationary
surface' in the sense that the evanescent functions are included.
We do so because the definition \eqn{approx} will be important further.
To characterize the
equivalence classes, we select from each one a representative function.
The set of representative functions will be denoted by ${\cal
F}^0_s\subset {\cal F}^0$.
In the example, this set can be defined in the following way.
We use the first field equation to write any function of $\bar \partial
\partial X$
as $\partial \left( h(\partial X^\mu )\right) $, so that the functions
${\cal F}^0_s$ depend on $X$, $\bar \partial^n X$, $\partial ^n X$, but
not on $\bar \partial ^m\partial ^n X$ for $m,n>0$. The second
field equation further restricts this set of functions.
It removes functions proportional to $\partial X^\mu
\cdot\partial X^\mu $.

Consider now any function $F\in {\cal F}^0$. Then we denote the
representative of
the class of functions which equals $F$ on the stationary surface by
$F_0\in {\cal F}^0_s$. We have
thus $F=F_0+y_i G^i$. The $G^i$ may be not unique. This is the case if
there are gauge invariances:
\begin{equation}
y_i R^i{}_{a}=0\ . \label{gaugeinv}
\end{equation}
Note that we use $a$, the index of the ghosts, for denoting all the
relations of this kind (we will give below a more exact definition). We
will see that we need a ghost
corresponding to each such relation. Anyway, we can build an expansion
\begin{equation}
F = F_0 +y_iF^i_1 +\ft12 y_iy_j F^{ji}_2 +\ldots \ .\label{Fexpy}
\end{equation}
The coefficient functions $F_0,F^i_1,F^{ij}_2,\ldots $ should
belong to the restricted set of the functions ${\cal F}^0_s$. This
expansion has of course graded symmetric coefficients. It
defines a derivative
\begin{equation}
\frac{\dl}{\partial y_i}F\equiv F^i_1 +y_j F^{ji}_2 +\ldots \ ,
\label{dFdy}\end{equation}
which is not uniquely defined if there are gauge invariances,
\eqn{gaugeinv}. Then the derivative of $F$
w.r.t. $y_i$ is only defined up to terms $R^i{}_{a}\epsilon ^a$, where
$\epsilon ^a$ is an arbitrary local function.
We demand there is no other indefiniteness in the definition of the
derivative $\partial F/\partial y_i$. This is equivalent to the statement
that our set of coefficient functions
$R^i{}_{a}$ be complete in the following sense: if for any set of local
functions $T^i(x)\in {\cal F}^0$ \begin{equation}
y_i T^i(x)=0\ \Longrightarrow \ T^i(x) = R^i{}_a \mu ^a(x)
+y_j v^{ji}(x) \label{allsym}\end{equation}
where $\mu ^a(x)$ and $v^{ij}(x)$ are local functions, the latter being
graded antisymmetric
\begin{equation}
v^{ij}(x)=(-)^{ij+1}v^{ji}(x) \ . \label{vantis}
\end{equation}
To prove this equivalence, one constructs $\mu ^a$ perturbatively in
powers of $y_i$.

The fields $\phi ^i$ are only a subset of the total set of fields $\Phi ^A$
and $\Phi ^*_A$, which we will collectively denote by $z^\alpha $. The
set
of functions of $\phi ^i$ denoted by ${\cal F}^0$ can be extended to
functions of $z^\alpha $. We will define the set ${\cal F}$ as polynomials
in the fields (and antifields) of ghost number different from zero, with
coefficients in ${\cal F}^0$. We have then
the following structure
\begin{equation}
{\cal F}\supset {\cal F}^0 \supset {\cal F}^0_s\ .
\end{equation}
Further we can define mappings
\begin{eqnarray}
P&:&{\cal F}\rightarrow {\cal F}^0\nonumber\\
 && F(z)\mapsto P(F) (\phi ^i)=F\left(\Phi ^*_A=0,\phi
^i,c^{a_k}=0\right)\nonumber\\
p&:&{\cal F}^0\rightarrow {\cal F}^0_s\nonumber\\
&& f(\phi ^i)\mapsto f_0(\phi ^i)\in {\cal F}^0_s \mbox{ and }f_0\approx f\
.\label{projections} \end{eqnarray}
For general local functions $F\in {\cal F}$, the symbol $\approx $
will be used in the following sense. These functions are defined as an
expansion in the fields of non--zero ghost number with coefficients in
${\cal F}^0$. Now one applies in each coefficient the relation
$\approx $ in ${\cal F}^0$. In other words, one uses the field
equations of $S^0(\phi )$ and leaves the fields of non--zero ghost
number as they are. So \eqn{approx} still applies.\vspace{1cm}

The master equation $(S,S)=0$ implies relations typical for a general gauge
theory. In the collective notation of fields and antifields, the master
equation takes the form
\begin{equation}
 S\dr_\alpha \cdot\omega ^{\alpha \beta }\cdot\dl_\beta S = 0 \ \mbox{
with }\ \omega ^{\alpha \beta }=(z^\alpha ,z^\beta )\ .\label{masteral}
\end{equation}
We also introduce the Hessian
\begin{equation}
Z_{\alpha \beta }\equiv P\left( \dl_\alpha  S\dr_\beta\right)\ .
\label{defZal}\end{equation}
Because $S$ has zero ghost number, $Z_{\alpha
\beta }$ is only non--zero if $gh(z^\alpha
)+gh(z^\beta )=0$. This implies that its non--zero elements are $
Z_{ij}= S^0_{ij}=\dl_i S^0\dr_j$, $Z^i{}_{a}$, $Z^a{}_{a_1}$, ...,
$Z^{a_k}{}_{a_{k+1}}$.
Note that upper indices of $Z$ appear here because derivatives are taken
w.r.t. antifields $\Phi ^*_A$. Of course also elements as
$Z_{a_{k+1}}{}^{a_k}$ are non--zero, being the supertransposed of the above
expressions.

{}From the ghost number requirements we can determine that $S$
is of the form
\begin{equation}
S=S^0+\phi ^*_i Z^i{}_{a}(\phi ) c^a+\sum
_{k=0}c^*_{a_k}Z^{a_k}{}_{a_{k+1}}c^{a_{k+1}}+\ldots \ ,\label{SinZ}
\end{equation}
where $\ldots$ stands for terms cubic or higher order in fields of
non--zero ghost number.

Considering the master equation at antifield number zero, we obtain
\begin{equation}
0=\ft12\left.(S,S)\right|_{\Phi ^*=0}=S^0\dr_i \cdot
\dl{}^i S^1=y_iZ^i{}_{a}c^a\ .
\end{equation}
These relations are the same as \eqn{gaugeinv}. The
transformation matrices $R^i{}_{a}$ can thus be taken to be $Z^i{}_{a}$.

Taking two derivatives of \eqn{masteral}, and applying the projection
$P$, we get
\begin{equation}
Z_{\alpha \gamma }\omega ^{\gamma \delta } Z_{\delta \beta }=(-)^{\alpha
+1}y_i \ P\left(\dl{}^i\dl_\alpha  S\dr_\delta\right) \ .  \end{equation}
This says that the Hessian $Z_{\alpha \beta }$ is weakly nilpotent.
Explicitly, we obtain (apart from the derivative of \eqn{gaugeinv})
\begin{eqnarray}
&&R^{i}{}_{a}Z^a{}_{a_1}=2 y_j f^{ji}{}_{a_1}
\approx 0 \label{RZisf}\\
&&Z^{a_k}{}_{a_{k+1}}Z^{a_{k+1}}{}_{a_{k+2}}\approx 0\ ,\label{ZZa0}
\end{eqnarray}
where
\begin{equation}
f^{ji}{}_{a_1}=\ft12(-)^i \ P\left( \dl{}^j \dl{}^iS\dr_{a_1}\right)\ .
\label{deffa1}\end{equation}
In the first relation the r.h.s. is written explicitly because it exhibits
a graded antisymmetry in $[ij]$.

A requirement on the extended action which was already briefly mentioned in
the introduction is the {\bf properness condition}. It concerns the
rank of
the Hessian. As the latter is weakly nilpotent its maximal (weak) rank is
half its dimension. The properness condition is now the requirement that
this matrix is
of maximal rank, which means that for any (local) function $v^\alpha (z)$
\begin{equation}
Z_{\alpha \beta }v^\beta \approx 0 \Longrightarrow v^\beta \approx \omega
^{\beta \gamma }Z_{\gamma \delta }w^\delta\ ,\label{proper}  \end{equation}
for a local function $w^\delta $.

The properness conditions \eqn{proper} can now be written explicitly as
\begin{eqnarray}
S^0_{ij}v^j\approx 0 &\Longrightarrow& v^j\approx
R^j{}_{a}w^a\nonumber\\
R^i{}_{a}v^a\approx 0 & \Longrightarrow& v^a\approx
Z^a{}_{a_1}w^{a_1}\nonumber\\
Z^{a_k}{}_{a_{k+1}}v^{a_{k+1}}\approx 0 &\Longrightarrow &
v_{a_{k+1}}\approx Z^{a_{k+1}}{}_{a_{k+2}}w^{a_{k+2}} \nonumber\\
&\mbox{and}&\nonumber\\
v_i R^i{}_{a}\approx 0 &\Longrightarrow& v_i \approx w^jS^0_{ji}\nonumber\\
v_a Z^a{}_{a_1}\approx 0 &\Longrightarrow& v_a\approx w_i
R^i{}_{a}\nonumber\\
v_{a_k} Z^{a_k}{}_{a_{k+1}}\approx 0 &\Longrightarrow& v_{a_k}\approx
w_{a_{k-1}} Z^{a_{k-1}}{}_{a_k} \ .
\label{properZ}\end{eqnarray}
The second group of equations follows from the first group, using that the
right and left ranks of matrices are equal. The first one implies
\eqn{allsym} if there are no non--trivial symmetries
which vanish at stationary surface. By the latter we mean that there would
be relations
\begin{equation}
y_i y_j T^{ji}=0 \ \mbox{ where }y_jT^{ji}\neq R^i{}_{a}\epsilon ^a
\end{equation}
and $T^{ij}$ is graded symmetric. If such non--trivial symmetries would
exist, then \eqn{allsym} is an extra requirement with $T^i$ replaced by
$y_j T^{ji}$.

\subsection{Antifield expansion and the Koszul--Tate differential}
\label{ss:afnKT}
An expansion according to antifield number is often very useful. E.g.
to construct the extended action, one can always solve the master
equation expanding in this way. To show the existence and uniqueness of
such perturbative solutions, the essential ingredient is the `Koszul--Tate'
(KT) differential.  It acts on antifields, is a nilpotent operation and has
an acyclicity property.  This means that its cohomology will consist only
of the functions on the stationary surface.

We start to define an expansion according to antifield number for
the antibracket operation:
\begin{eqnarray}
(F,G) &=& \sum_{k=1} (F,G)_k \nonumber\\
(F,G)_1&=&\sum_{i}( F\dr_{i}\cdot \dl{}^{i} G - F\dr{}^i\cdot  \dl_i
G)\nonumber\\
(F,G)_{k+2}&=&\sum_{a_k}( F\dr_{a_k}\cdot  \dl{}^{a_k} G -F\dr{}^{a_k}
\cdot \dl_{a_k}G)\ .
\end{eqnarray}
Note that the last sum is taken with the fixed value of $k$, but over all
values of $a_k$. The subindex is chosen such that
\begin{equation}
afn \left( (F,G)_k\right)  = afn(F) + afn(G) -k\ .
\end{equation}

We will often have to consider functions $F(\Phi ,\Phi ^*)$ satisfying
\begin{equation}
{\cal S}F\equiv (F,S)=\sum_{k,\ell,m} (F^k,S^\ell )_m=0\
,\label{expSF} \end{equation}
where we expanded as well $F$, $S$ as the antibracket in antifield
numbers.
Let us denote the ghost number of $F$ by $f$. If $f$ is negative, then
all terms of $F$ should contain antifields, and the sum over $k$
starts at $k=-f$.
Otherwise this sum, as that over $\ell $, starts at $k,\ell =0$
(while the sum over $m$ starts at $m=1$).
Concerning the ghosts it is useful to introduce a `pureghost' number, which
is equal to the ghost number for fields, and zero for antifields. We thus
have
\begin{eqnarray}
&&puregh(c ^{a_k})=k+1\ ;\qquad puregh(\phi ^*_{a_k})=0 \nonumber\\
&&gh\ =\ puregh\ -\ afn\ .
\end{eqnarray}
For the ranges of $k,\ell$ and $m$ in the sum \eqn{expSF} we can make
restrictions
using that the pureghost number of $F^k$ is $f+k$, and in the term
$(\cdot,\cdot)_m$ there are derivatives w.r.t. antifields of antifield
number $m$ and w.r.t. fields of pureghost number $m-1$. So we have
respectively  if $F^k$ is derived w.r.t. a field and $S^\ell$
w.r.t. an antifield, or in opposite order
\begin{equation}\begin{array}{ll}
m-1\leq f+k \ ;\qquad& m\leq \ell\ ,\\ \hspace{3cm} \mbox{or}&\\
m\leq k \ ;\qquad& m-1\leq  \ell   \ .
\end{array}
\end{equation}
We collect the terms with equal value of the antifield number of
${\cal S}F$, so $k+\ell-m=n$.  According to the above inequalities,
there is then just one case where $\ell-m$ is negative, so where the
antifield number of ${\cal S}F$ is lower than that of $F$.  We split
this term from the others and obtain
\begin{eqnarray}
\left( {\cal S}F\right) ^n&=&(-)^F\dkt  F^{n+1}
+D^nF(S^1,\ldots S^{\tilde n},F^0, \ldots, F^n)  \label{cSFexp}\\
\dkt F &=& \sum_{k=0}(S^k,F)_{k+1}=
\sum_{k=-1}S^{k+1}\dr_{a_k}\cdot\dl{}^{a_k}F \ , \label{defdkt0}\\
D^nF&\equiv&
\sum_{k=0}^{n} \sum_{m=1}^{\tilde k}(F^k,S^{n-k+m})_{m}\
,\label{defDn} \end{eqnarray}
where $\tilde k=k$ if $f< 0$ and $\tilde k =k+f+1$ for $f\geq 0$.
The second line defines the Koszul--Tate differential. We have chosen
it to act from the left in accordance with previous references. Note
the following important facts about it. First it is a fermionic
operation, which means that $\dkt F$ has opposite statistics as $F$.
It lowers the antifield number by one, and it acts only non--trivially
on antifields. Using \eqn{SinZ} we find
\begin{eqnarray}
\dkt \phi ^*_i &=& y_i\nonumber\\
\dkt c^*_a &=& \phi ^*_i R^i{}_a\nonumber\\
\dkt c^*_{a_k}&=&c ^*_{a_{k-1}}Z^{a_{k-1}}{}_{a_k}+M_{a_k}(\phi
;\phi ^*_i,c^*_a,\ldots ,c^*_{a_{k-2}})\mbox{   with }k\geq 1 \ ,
\label{defdkt}
\end{eqnarray}
where $M_{a_k}(\phi ;\phi ^*_i,c^*_a,\ldots ,c^*_{a_{k-2}})$ is determined
by the parts not explicitly written in \eqn{SinZ}.

The important properties of this operation are its nilpotency and
acyclicity. These are the remnants of respectively the master equation
and the properness condition of the extended action. First, the
nilpotency
is easy to show when we already know that $(S,S)=0$ \footnote{The
KT differential is introduced e.g. in \cite{Henn} in a different way
to prove that there
is a solution to the master equation. Then the proof of the
nilpotency is more involved.}. In fact on functions of a definite
antifield number we have
\begin{eqnarray}
{\cal S}F^k&=&(-)^F\dkt F^k+\sum_{n\geq k}D^n F^k\nonumber\\
\dkt F^k&=& (S,F^k)^{k-1}\ ,\label{cSFk}
\end{eqnarray}
where the superscript means that we restrict to the terms of antifield
number $k-1$ (and thus omit the terms of higher antifield number).
Then
\begin{equation}
\dkt\dkt F^k=\left( S,(S,F^k)^{k-1}\right) ^{k-2}=\left(
S,(S,F^k)\right) ^{k-2}=0\ . \end{equation}
Indeed, we can omit the superscript of the first antibracket, as the other
terms have after the two antibracket operations an antifield number which
is higher than $k-2$. Then
we use the Jacobi--identities \eqn{Jacobi} and the master equation.
Therefore we can define a cohomology.

\subsection{Acyclicity of the KT differential.}\label{ss:KTacycl}
A more difficult, but essential, property of the KT differential is
its acyclicity. As the operation only acts on antifields, we now have
to consider the function
space ${\cal F}^*\equiv \{F(\phi^i ,\Phi ^*_A)\}\subset {\cal F}$. The
acyclicity statement is that the cohomology of the KT differential on
this space contains only the functions on the stationary surface ${\cal
F}^0_s$. \begin{equation}
\dkt F(\phi ,\Phi ^*) =0\ \Rightarrow \ F=\dkt H + f(\phi ) \label{acycl}
\end{equation}
where $f\in {\cal F}^0_s$.

The proof of this statement goes perturbatively in the level of antifields
which are included in the set. We define ${\cal F}^*_k=\{F(\phi ^i,\phi
^*_i,\ldots ,c^*_{a_{k-2}}) \}$ as the functions which depend at most
on antifields with antifield number $k$, such that
\begin{equation}
{\cal F}\supset{\cal F}^*\supset \ldots \supset {\cal F}^*_k\supset \ldots
\supset {\cal F}^*_1 \supset {\cal F}^*_0\equiv {\cal F}^0 \supset {\cal
F}^0_s \ .\end{equation}
In ${\cal F}^*_0={\cal F}^0$ we have that the functions which should
not be in the cohomology according to \eqn{acycl} are the functions
which are proportional to a field equation $y_i$. This is $\dkt \phi
^*_i$. So we need functions in ${\cal F}^*_1$ to remove these.
Lemma~\ref{lem:dkt0r} in the appendix~\ref{app:lemprKT} proves the
acyclicity
in ${\cal F}^*_1$ apart from terms proportional to $\phi^*_i
R^i{}_{a}=\dkt c^*_a$. Then one proves a similar lemma in ${\cal
F}^*_2$, where now only terms proportional to $\dkt c^*_{a_1}$ are
left in the cohomology (apart from the functions on the stationary
surface). Continuing in this way we can perturbatively construct the
function $H$ in \eqn{acycl}. The essential ingredient for all these
proofs is of course the properness condition. It is this condition
which is translated to the acyclicity of the KT differential.

The proofs are in appendix~\ref{app:lemprKT} because
they are rather long and tedious. Although they are related to the
proofs of \cite{Henn} they are fundamentally different, as was explained in
the beginning of section~\ref{s:adbkta}.

In
this appendix~\ref{app:lemprKT} we first show the relation between the
KT differential
in different sets of coordinates, and prove that if acyclicity is true
in one set of coordinates, then it is also true in a canonically
related set. Of course, these canonical transformations do not include
the interchange of fields and antifields, as
canonical transformations preserve the ghost number and we always have to
reserve
the word `antifields' here for those fields of negative ghost
number. Then the proofs are given for the first levels, such that the
generalization is obvious. It shows that perturbatively
in the functions containing antifields of higher antifield numbers,
the KT differential contains no non--trivial cycles apart from the
(antifield independent) functions at ghost number zero in ${\cal
F}^0_s$ which are the functions on the stationary surface.

This acyclicity statement applies for local ($x$--dependent) functions. It
does not apply in general for integrals. Indeed, consider the following
example \begin{eqnarray}
S&=&\int d^dx\, \ft12\partial _\alpha  X^\mu \cdot \partial ^\alpha  X^\mu
\ ,\ \mbox{ with }\mu =1,2\  \  \alpha =1,...,d\nonumber\\
F&=&\int d^dx\, \left( X^*_1 X^2 -X^*_2 X^1\right)\ \rightarrow \ \dkt F=0\
. \end{eqnarray}
Nevertheless, $F$ can not be written as $\dkt G$. This violation of
acyclicity for integrals is due to
rigid symmetries, such that \eqn{allsym} does not apply for integrals.
However, for $F$ a local integral (integral of a local function) of
antifield number 0 (and thus obviously $\dkt F=0$),
which vanishes on the stationary surface, we have by definition $F=\int y_i
F^i$ and $F=\dkt\int (\phi ^*_i F^i)$.

The acyclicity was proven here for functions independent of ghosts
(elements of ${\cal F}^*$).
If one considers functions $F(\Phi ^*,\phi ,c)\in {\cal F}$ depending
on ghosts, then the acyclicity can be used when we first expand in
$c$. Therefore the modified acyclicity statement is then
\begin{eqnarray}
\dkt F(\Phi ^*,\phi ,c) =0 &\Rightarrow & F=\dkt H + f(\phi,c
)\nonumber\\ \mbox{and if }f(\phi ,c)\approx 0&\Rightarrow & f=\dkt
G(\Phi ^*,\phi ,c)\ , \label{acyclc} \end{eqnarray}
where, as mentioned before, $\approx $ stands for using the field
equations of $S^0(\phi
)$ for $\phi $, while the ghosts $c$ remain unchanged.
If $F$ is an integral, where the
integrand contains ghosts, then we apply the acyclicity to the
coefficient functions of the ghosts which are local functions. If
$gh(F)>0$ (or even if just $puregh(F)>0$) then each term can be treated in
this way and the statement
\eqn{acyclc} holds even when $F$ is an integral of a local quantity. Also
if $gh(F)=0$ then any term has either a ghost, or it just depends on $\phi
^i$ in which case the acyclicity statement also applies for integrals.

\subsection{Antibracket cohomology}
\label{ss:ABcoho}
In this section we present our main theorem, mentioned in the introduction.
We consider the problem that we have to find a certain function
$F(\Phi ,\Phi ^*)$, which is known to satisfy $(F,S)=0$. We want to know
$F$ modulo a part $F=(G,S)$. We thus investigate here the antibracket
cohomology.

First we have to consider which functions are invariant under the
${\cal S}$ operation: see \eqn{cSFexp}, and \eqn{cSFk}. Especially the
last expression of ${\cal S}$ is useful, as it shows that the equation
at lowest antifield number of ${\cal S}F^k=0$ is $\dkt F^k=0$ ($k\geq 1$).
This is the clue for solving ${\cal S}F=0$ perturbatively in antifield
numbers.

First, in theorem~\ref{thm:ABcohoneg} it is proven that there is no
cohomology for local functions of negative ghost number.
For  non--negative ghost numbers the situation is more complicated. The
equation ${\cal S}F=0$ at zero antifield number is by \eqn{cSFexp}
\begin{equation}
-(-)^F\dkt F^1 =D^0F^0=\sum_{m=1}^{f+1} (F^0,S^m)_m \ .
\label{cSF00} \end{equation}
$D^0 $ is a fermionic right derivative operator,
which acts on fields only, and is given by
\begin{equation}
D^0  F^0  =
\left. (F^0,S)\right|_{\Phi ^*=0}
\ .\end{equation}
For antifield number 0, the KT differential is
acyclic on functions which vanish on the stationary surface.
Therefore $F^1$ exists if $D^0 F^0\approx 0$, and in lemma~\ref{lem:F0F1SF}
it is proven that then also the full $F$ can be constructed perturbatively
in antifield number such that ${\cal S}F$=0.

This operator $D^0 $ raises the pureghost number by 1. It is nilpotent on
the classical stationary surface: $D^0D^0 F^0\approx 0$. One can define
a (weak) cohomology of this operator (lemma~\ref{lem:nilpD0}) on functions
of fields only, and this is graded by the pureghost
number $p$. The main result is that this cohomology is equivalent to the
(strong) cohomology of ${\cal S}$ for
functions of ghost number $p$ (theorem~\ref{thm:cohoabD0}). This is
the theorem which determines the antibracket cohomology. We can state
it as follows
\begin{theorem} : Any local function $F$ of negative ghost number which
satisfies ${\cal S}F=0$ can be written as $F={\cal S}G$. \\
For functions of non--negative ghost number, the following statement holds.
For a local function or a local integral $F^0(\Phi )$ (not containing
antifields)
\begin{equation}
D^0 F^0\approx 0\ \Leftrightarrow  \ \exists F(\Phi,\Phi ^*)\ : {\cal
S}F=0 \label{mainth1}\end{equation}
where $F(\Phi ,0)=F^0$. Further
\begin{equation}
F^0\approx D^0G^0\  \Leftrightarrow \ \exists G(\Phi ,\Phi ^*)\ : F={\cal
S}G\ ,
\end{equation}
where again $G^0=G(\Phi ,0)$, and $F$ is a function determined by
\eqn{mainth1}. The ghost numbers of $F$ and $G$ are equal to the
pureghost numbers of $F^0$ and $G^0$. \label{thm:main}
\end{theorem}
The inclusion of local integrals for the second part of the theorem follows
from the fact that for non--negative ghost numbers we could at the end of
subsection~\ref{ss:KTacycl} include these in the acyclicity statement, and
this was the only ingredient of the proof.

At ghost number zero the antibracket cohomology gives the
functions on the stationary surface, where two such functions
which differ by gauge transformations are
identified . Indeed, the KT cohomology
reduced the functions to those on the stationary surface.
$D^0$ acts within the stationary surface, and its cohomology reduces
these functions to the gauge invariant ones, as
for (pure) ghost number 0,
\begin{equation}
D^0  F^0=F_0\dr_i \cdot R^i{}_{a} c^a\ ,
\end{equation}
gives the gauge transformation of $F^0$. Here we clearly see how
the antibracket and BRST formalisms are connected. $D^0$ is in fact
the BRST operator.

At ghost number 1, we can apply the theorem for the analysis of anomalies.
We have seen in section~\ref{ss:ingrBV}
that anomalies ${\cal A}$ are local integrals of ghost number 1. They
satisfy the Wess--Zumino consistency relations \cite{WZcc} in the
form\footnote{We consider here only 1--loop effects.} ${\cal
S}{\cal A}=0$, while anomalies can be absorbed in local counterterms if
${\cal A}={\cal S}M$. The anomalies are thus in fact elements of
the cohomology of ${\cal S}$ at ghost number 1 in the set of local
integrals. We have found here that in the classical basis,
these anomalies are completely determined by their part ${\cal
A}^0$ which is independent of antifields and just contains 1 ghost of ghost
number 1. On this part there is the consistency condition $D^0 {\cal
A}^0\approx 0$.  If this equality is strong, then ${\cal A}$ does not need
antifield--dependent terms for its consistency. If it is weak, then
these
are necessary, but we know that they exist. If for an anomaly ${\cal
A}^0\approx D^0 M^0$, then we know that it can be cancelled by a local
counterterm. Consequently the
anomalies (as elements of the cohomology) are determined by their
part without antifields, and with ghost number one, and can thus be
written as
\begin{equation}
{\cal A}={\cal A}_a (\phi )c^a+\ldots\ ,
\end{equation}
where the written part determines the $\ldots$.
Therefore we can thus
split the anomalies in parts corresponding to the different symmetries
represented by the index $a$.
Indeed, people usually talk about anomalies in a certain symmetry (although
this can still have different forms according to the particular
representant of the cohomological element which one considers), and we show
here that this terminology can always be maintained for the general gauge
theories which the BV formalism can describe.

\section{Consistent Anomalies in $W_3$ gravity}\label{ss:W3}
\setcounter{equation}{0}
\subsection{The classical theory and the extended action}
The classical action for chiral $W_3$ gravity is \cite{HULL}\footnote{We
will use the notations $\partial =\partial _+$ and $\bar \partial
=\partial _-$, where $x^\pm=\rho (x^1\pm x^0)$, and we leave the factor
$\rho $ undetermined.\label{fn:convrho}}${}^,$ \footnote{We omit the
$\int d^2 x$ in all the actions below. Similarly for matrices as $S_{AB}$
there is a factor $\delta (x-y)$ if the DeWitt index $A$ contains the point
$x$, and $B$ the point $y$. Derivatives which will still act to the right
in these matrices (all those not enclosed in brackets) act on these delta
functions. We will not write the latter explicitly.}
\begin{equation}
S^0=-\ft{1}{2}\pl \bpl +\ft{1}{2}h\pl
\pl +\ft{1}{3}\ds B\pl\plb \plg \,\label{S0W3} \end{equation}
where $\ds$ is a symmetric tensor satisfying the nonlinear identity
\begin{equation}
d_{\mu (\nu \rho }d_{\sigma )\tau  \mu }=\kappa \delta _{(\nu
\sigma }\delta _{\rho )\tau  }\,\label{dsym} \end{equation}
for some arbitrary, but fixed, parameter $\kappa $. The general solution of
this equation was found in \cite{dsymb}\footnote{The general solutions
are related to specific realizations of real
Clifford algebras ${\cal C}(D,0)$ of positive signature. We have then
$n=1+D+r$, where $r$ is the dimension of the Clifford algebra realization.
We obtain solutions of \eqn{dsym} for the following values :
$(D=0,r=0)$, $(D=1,r$ arbitrary), $(D=2,r=2)$, $(D=3,r=4)$ (these are the
$SU(3)$ $d$--symbols), $(D=5,r=8)$,
$(D=9,r=16)$. The latter four are the so--called `magical cases'. Then the
Clifford algebra representation is irreducible, and the $d$--symbols are
traceless as it is the case for $(D=1,r=0)$. All the solutions can be
given in the following way.
$\mu$ takes the values $1,a$ or $i$, where $a$ runs over $D$ values and $i$
over $r$ values. The non--zero coefficients are (for $\kappa =1$)
$d_{111}=1$, $d_{1ab}=-\delta _{ab}$, $d_{1ij}=\frac{1}{2}\delta _{ij}$,
and $d_{aij}=\frac{\sqrt{3}}{2}(\gamma _a)_{ij}$. For $D=1$ the gamma
matrix is $(\gamma _a)_{ij}=\delta _{ij}$ (reducible). In that case the
form of the solution can be simplified by a rotation between the index 1,
and $a$, which takes only one value, see below: \eqn{solD1}. All
representations
of all real Clifford algebras ${\cal C}(D,0)$ appear as solutions of a
generalization of \eqn{dsym} and classify the homogeneous special
K\"ahler and quaternionic spaces \cite{alhom}\label{fn:soldsym}}.
The model contains $n$ scalar fields $X ^\mu
, \mu =1,...,n$ and two gauge fields $h$ and $B$.  The field equations are
\begin{eqnarray}
y_\mu &=&\partial \bar{\partial }X^\mu -\partial
(h\pl)-\ds\partial (B\plb \plg)\nonumber\\
y_h&=&\ft{1}{2}\pl \pl\nonumber\\
y_B&=&\ft{1}{3}\ds \pl \plb \plg\,.\label{yW3}
\end{eqnarray}
As one can check, by using \eqn{dsym}, these field equations are not
independent,
but satisfy
\begin{equation}
y_iR^i{}_{a}=0\,\,\,\, i=\{\mu ,h,B\} \,\,\,\, a=1,2\,. \label{inv}
\end{equation}
It expresses that the action \eqn{S0W3} has two gauge invariances $\delta
\phi ^i=R^i{}_{a}\epsilon ^a$, corresponding to
the local reparameterization and $W_3$ symmetry. Therefore, we
introduce the ghosts $c$ and $u$, and
$R^i{}_{a}$ is given by
\begin{eqnarray}
S^1\ =\ \phi ^*_iR^i{}_ac^a&=&
X _\mu ^*\left[\pl c+\ds \plb \plg u\right]\nonumber\\
& &+h^*\left[(\na ^{-1}c)+\ft{\kappa }{2}\pl \pl
(D^{-2}u)\right]\nonumber\\ &&+B^*\left[(D^{-1}c)+(\na ^{-2}u)\right]
\label{S1W3}\end{eqnarray}
and
\begin{eqnarray}
\na ^j=\bar{\partial }-h\partial -j(\partial h)\nonumber\\
D^j=-2B\partial -j(\partial B)\,.
\end{eqnarray}
The number $j$ is the spin (the number of unwritten $+$ indices).
These spins are given in table~\ref{tbl:fieldsW3} together with other
properties of derivative operators and fields.
\begin{table}[htf]
\label{tbl:fieldsW3}\begin{center}\begin{tabular}{||l|r|r|r||l|r|r|r||}
\hline
 & $gh$& $j$    & $dim-j$&        &$gh$ &$j$ &$dim-j$ \\ \hline
$X$& 0   & 0      & 0      & $X^*$  &  $-1$& 0  & 2 \\
$B$& 0   &$-3$    & 2      & $B^*=v$&  $-1$&$3$ & 0 \\
$h$& 0   &$-2$    & 2      & $h^*=b$&  $-1$&$2$ & 0 \\
$c$& 1   &$-1$    & 0      & $c^*$  &  $-2$&$1$ & 2 \\
$u$& 1   &$-2$    & 0      & $u^*$  &  $-2$&$2$ & 2 \\ \hline
$\partial$ &0  & 1    & 0 &&&& \\
$\bar \partial,\,\nabla$  & 0 &$-1$& 2    &&&& \\
$D$ & 0 &       $-2$     & 2    &&&& \\ \hline
\end{tabular}\end{center}
\caption{We give here the properties of fields and derivative operators.
$gh$ is the ghost number and $j$ is the spin. Further
one can assign an `engineering' dimension to the fields and derivatives,
such that the Lagrangian has dimension 2. We defined $dim(\Phi
^*)=2-dim(\Phi )$ (the number 2 is arbitrary, this freedom is
related to redefinitions proportional to the ghost number). When we
subtract the spin from this dimension,
we find that only a few fields are dimensionfull. We mentioned already
the new names of fields in the gauge fixed theory}. \end{table}

The matrix $R^i{}_{a}$
is of maximal rank, so there are no further zero modes, i.e.
$Z^{a_k}{}_{a_{k+1}}=0, k\geq 0$ and we have no further ghosts. The
calculation of $S^2$ amounts to calculate the structure functions $T$ and
nonclosure functions $E$ of the gauge algebra:
\begin{equation}
S^2=(-)^b\ft{1}{2}c^*_aT^a{}_{bc}c^cc^b
+(-)^{i+a}\ft{1}{4}\phi ^*_i\phi ^*_jE^{ji}{}_{ab}c^bc^a\ .
\label{S2form}
\end{equation}
After calculating $(S^1,S^1)$ we see that we can choose the structure
functions $T^a{}_{bc}$ either such that this term in $(S^2,S^1)$ cancels
all terms with $X^*_\mu $, or such that all terms with $B^*$ are
cancelled. So one can choose which of the two antifields appear in $S^2$.
In other words, in the first case the algebra is chosen to close off--shell
on $X^\mu $ and to close only on--shell on $B$, while this is reversed
in the second case. In fact one can give the following expression for $S^2$
which depends on a free parameter $\alpha $
\begin{eqnarray}
S^2 &= &c^*\left[(\partial c)c+\kappa (1-\alpha ) \pl \pl (\partial
u)u\right]+u^*\left[2(\partial c)u-c(\partial u)\right]\nonumber\\
& &-2\kappa\alpha  h^*(D^3B^*+\na^2h^*)(\partial u)u -
2\kappa(\alpha+1)  X_\mu ^*h^*(\partial u)u\pl \ .
\label{S2W3}
\end{eqnarray}
$\alpha =-1$ corresponds to the first choice discussed above, while
$\alpha =0$ corresponds to the second choice, and this form has been
written in \cite{Hullmatter,SSvNMiami}. Remark also the choice
$\alpha =1$ which gives the simplest structure functions. The relation
between these
actions is given by the canonical transformation: starting from the action
with $\alpha =0$ the transformation with generating function
\begin{equation} f=2\kappa\alpha  h'^*(\partial u)uc'^*\,.
\label{cantrlambda}\end{equation}
gives in the primed coordinates the action with this arbitrary parameter
$\alpha $. This (fermionic) generating function $f(\Phi ,\Phi '^*)$
determines a canonical transformation by \cite{BVcan,anombv,bvleuv}
\begin{equation}
\Phi '^A=\Phi ^A +\frac{\partial }{\partial \Phi '^*_A}f(\Phi ,\Phi '^*)\
;\qquad
\Phi ^*_A=\Phi '^*_A+\frac{\partial }{\partial \Phi ^A}f(\Phi ,\Phi '^*)\ .
\label{cantrf}\end{equation}
Note that this type of freedom is always possible. A canonical
transformation with $f=c'^*_a\phi '^*_i T^{ia}{}_{bc}c^cc^b (-)^{a+b}$ adds
a term with a field equation in the structure functions and changes the
non--closure functions. In general this can also change the higher
antifield
terms of the extended action. In the case at hand however, the master
equation is already satisfied for all $\alpha $ without the need of a term
$S^3$. This is equivalent to the statement that the Jacobi identities
\begin{equation}
(-)^b\left( T^a{}_{bc}c^cc^b\right) \dr_i\cdot
R^i{}_d\,c^d+(-)^{b+d}T^a{}_{bc}\,c^c\,T^b{}_{de}\,c^e\,c^d\approx 0
\end{equation}
are strongly satisfied (with $=$ rather than $\approx $). The canonical
transformation \eqn{cantrlambda} does not generate terms in $S$ of
antifield number 3 or higher. So the full extended action is
\begin{equation}
S=S^0+S^1+S^2\ ,
\end{equation}
where the three terms are given  in \eqn{S0W3}, \eqn{S1W3} and
\eqn{S2W3} respectively.

Now, in order to obtain a gauge fixed action, we perform the
canonical transformation from $\{h,h^*,B,B^*\}$ (fields and antifields of
the classical basis) to new fields and antifields $\{b,b^*,v,v^*\}$ (the
gauge--fixed basis):
\begin{equation}
h=-b^*;\ h^*=b\ \mbox{  and  }\ B=-v^*;\ B^*=v\,.\label{ctrW3gf}
\end{equation}
One may check now that the new antifield independent action, which depends
thus on $\{X^\mu ,b,c,v,u\}$, has no gauge invariances. These are thus
the fields that appear in loops.

\subsection{Calculation of the one--loop anomaly without antifields}
\label{ss:calc1lW3}
Now, we will calculate the one--loop anomaly following the
prescription summarized in section~\ref{ss:oneloop}. We need first
the (invertible) matrix of second derivatives w.r.t. the fields of the
gauge--fixed basis. The theorems proven in section~\ref{ss:ABcoho}
will imply that we will only need in these second derivatives the terms
without
fields of negative ghost numbers and at most linear in $c$ and $u$,
the fields of ghost number 1. We will therefore in the entries still use
the names of fields and antifields as in the classical basis.
In the following matrices we first write the
entries corresponding to the bosons $X^\mu $, and then order the fermions
according to ghost number and spin:
$\Phi '^A=\{X^\mu , v=B^*, b=h^*, c,u \}$ we have  $S'_{AB}=\tilde
S'_{AB}+$
terms of antifield number non--zero $+$ terms of pureghost number $>1$.
\begin{eqnarray}
\tilde S'_{AB}&=&\pmatrix{S_{\mu \nu }& q_\mu \cr -q_\nu ^T & \tilde \nabla
\cr}\nonumber\\[0.4cm]
S_{\mu \nu }&=&  \delta
_{\mu \nu }\nabla ^1\partial +d_{\mu \nu \rho }D^2\plg\partial
\nonumber\\
q_\mu &=&\pmatrix{ 0 & \kappa \partial (D^{-2}u)\pl &0 &0 \cr }\nonumber\\
\tilde \nabla &=&\pmatrix{ 0&0& D^{-1}&\nabla ^{-2}\cr
 0&0  & \nabla ^{-1}& \kappa y_h D^{-2}\cr
 D^3& \nabla ^2&0&0\cr
\nabla^3 & \kappa D^4y_h &0&0\cr }
\ , \label{SABgfW3}\end{eqnarray}
where $y_h$ is given by \eqn{yW3}.
A lot of zeros follow already from the ghost number requirements.

To obtain the regulator we have, according to \eqn{regulator}, to choose a
mass matrix $T$ and multiply its inverse with $\tilde S'$.  Looking at the
table~\ref{tbl:fieldsW3} for the spins of the fields, we notice that some
fermionic fields (e.g.  $c$) do not have a partner of opposite spin, which
we need for a mass term, if we want to preserve these rigid symmetries (the
PV partners of fields have the same properties as the corresponding
fields).  Further, the regulator will not regularize \eqn{BOXS} because the
fermion sector of \eqn{SABgfW3} is only linear in the derivatives.  These
two problems can be solved by first introducing extra PV fields (this
procedure was already used in \cite{anombv}).
They have no interaction in the massless sector and do not
transform under any gauge transformation. Inspecting the ghost numbers and
spins of the fermions in $\{\Phi '^A\}$, we find that we need extra PV
fields with ghost numbers and spin as in table~\ref{tbl:extraPVW3}.
\begin{table}[htf]
\label{tbl:extraPVW3}\begin{center}\begin{tabular}{||l|r|r|r||}
\hline
 & $gh$& $j$    & $dim-j$\\ \hline
$\bar u$& 1   &$-3$    & 1       \\
$\bar c$& 1   &$-2$    & 1       \\
$\bar b$& $-1$&$1$     & 1       \\
$\bar v$& $-1$&$2$     & 1       \\ \hline
\end{tabular}\end{center}
\caption{The extra non--interacting and gauge invariant PV fields. The
ghost
numbers and spins are chosen in order to be able to construct mass terms
for $c,u,b$ and $v$. The names are chosen such that $gh(\bar x)=gh(x)$
and $j(\bar x)=j(x)-1$. The dimensions follow from the kinetic terms,
although this would still allow less symmetric choices.} \end{table}
Then we have to choose the matrix $T$. It determines the mass matrix in the
PV action as in \eqn{SMPV}. We can choose it as
\begin{equation}
T =\pmatrix {\delta _{\mu \nu }& 0 & 0\cr 0 & 0&\frac{1}{M} \unity \cr
0&-\frac{1}{M}\unity &0\cr}\ , \end{equation}
where the second entry refers to the fermions from above, and the third
line to the
four new fermions. The latter are thus ordered as in the table. The kinetic
part of their action can be chosen such that the
enlarged matrix $S'_{AB}$ is \begin{equation}
S'_{AB}=\pmatrix{S_{\mu \nu }& q_\mu &0\cr -q_\nu ^T&\tilde \nabla &0\cr
0&0&-\tilde \partial \cr}\ ;\qquad
\tilde \partial \equiv \pmatrix {0 &0 & 0&\partial \cr 0 & 0 &
\partial & 0\cr 0&\partial &0&0\cr \partial &0&0&0\cr}\ .
\label{kinexPV} \end{equation}
To put everything together, we find as regulator
\begin{eqnarray}
&&{\cal R}={\cal R}_0 + M {\cal R}_1\nonumber\\
&&{\cal R}_0=\pmatrix{
S_{\mu \nu }& q_\mu &0\cr 0 & 0 &0 \cr 0 &0 &0\cr}\ ;\qquad
{\cal R}_1=\pmatrix{0& 0 &0\cr 0 & 0 &\tilde \partial \cr
-q^T_\nu &\tilde \nabla &0\cr}\ ,
\end{eqnarray}
and the transformation matrix $K$ defined in \eqn{defK} is, dropping again
terms of pureghost number 2 or antifield number 1,
\begin{eqnarray}
K&=&\pmatrix{K^\mu {}_{\nu }& K^\mu {}_{F}&0\cr
           K^F{}_{\nu }& K^F{}_{F}& 0 \cr
           0 & 0 & 0 \cr}\nonumber\\[0.4cm]
K^\mu {}_{\nu }&=&\delta ^\mu _\nu c\partial +2ud_{\mu \nu \rho }(\partial
X^\rho )\partial \nonumber\\
K^\mu {}_{F}&=&\pmatrix{0&0&\pl&d_{\mu \rho \sigma }(\partial X^\rho
)(\partial X^\sigma )\cr}\nonumber\\
K^F{}_{\nu }&=&\pmatrix{-d_{\nu \rho\sigma  }(\partial X^\rho )(\partial
X^\sigma )\partial\cr -(\partial X_\nu )\partial\cr 0\cr 0\cr}\\
K^F{}_{F}&=&\pmatrix{
- (c\partial )_{3} & -2\kappa[ y_h (u\partial )_2+u(\partial y_h)]&0&0\cr
-2(u\partial )_{\frac{3}{2}}&-(c\partial )_2&0&0 \cr
0&0&-(c\partial )_{-1}&-2\kappa (1-\alpha )y_h(u\partial )_{-1}\cr
0&0&-2(u\partial )_{-\frac{1}{2}}&-(c\partial )_{-2}\cr }\nonumber\ ,
\end{eqnarray}
where we used the shorthand
\begin{equation}
(c\partial )_{x}=c\partial +x(\partial c)\ .
\end{equation}
Note that the second line of \eqn{S2W3} can not play a role for these
expressions up to the order in which we have to calculate.

The expression of ${\cal R}$ is so far still linear in derivatives for the
fermionic sector. This we solve as in \cite{anomPV} by multiplying in
\eqn{SSMPV} the numerator and the denominator by $(1+{\cal R}_1/M)$. We
obtain
\begin{equation}
(\Delta S)^0=Tr\left[K(1+\frac{1}{M}{\cal R}_1)\frac{1}{\left(
1-\frac{1}{M^2}{\cal R}_0-\frac{1}{M}{\cal R}_1\right) \left(
1+\frac{1}{M}{\cal R}_1\right) }\right] \ , \end{equation}
which leads to the regulator
\begin{equation}
{\cal R}_0 +{\cal R}_1^2+\frac{1}{M}{\cal R}_0{\cal R}_1
=\pmatrix{ S_{\mu \nu }& q_\mu &\frac{1}{M}q_\mu \tilde \partial \cr
-\tilde \partial q^T_\nu & \tilde \partial \tilde \nabla & 0 \cr
 0 &0 &\tilde \nabla\tilde \partial \cr}\ .
\end{equation}
On the other hand
\begin{equation}
K{\cal R}_1 =\pmatrix{0&0&K^\mu {}_{F}\tilde \partial \cr
0&0&K^F{}_{F}\tilde \partial \cr 0&0&0\cr}\ .
\end{equation}
This can not contribute to the trace because grouping the first two
entries together, this regulator is also upper triangular.
Therefore we can omit the term $K{\cal R}_1$,
and then only the first two rows and columns of the
regulator can play a role. This eliminates also the $(1/M)$ terms in the
regulator. So we effectively have to calculate
\begin{equation}
(\Delta S)^0=Tr\left[\pmatrix{K^\mu {}_{\nu }& K^\mu {}_{F}\cr
           K^F{}_{\nu }& K^F{}_{F} \cr}\exp \frac{1}{M^2}\tilde {\cal R}
\right] \ ;\qquad {\cal R}= \pmatrix{ S_{\mu \nu }& q_\mu \cr
 -\tilde \partial q^T_\nu & \tilde \partial \tilde \nabla\cr} \ .
\end{equation}

As explained in section~\ref{ss:oneloop} such calculations can be done in
the heat kernel method \cite{heatk}, was it not that our regulator contains
terms with second derivatives which are not proportional to the unit matrix
in internal space.  We can anticipate on the form of the anomaly which we
have to obtain. The anomaly $\Delta S$ should be an integral of a local
quantity of spin 0 and dimension 2. From the $dim-j$ column of
table~\ref{tbl:fieldsW3}, we then see that the anomaly can be split in a
part without $B$, and a linear part in $B$, which can not contain $h$:
\begin{equation}
(\Delta S)^0=\Delta ^hS +\Delta ^BS\ .
\end{equation}
So we split $\tilde {\cal R}={\cal R}^h+{\cal R}_1$
\begin{eqnarray}
{\cal R}^h&=&diag\left( \delta_{\mu \nu }\partial \nabla ^0,\partial
\nabla
^3,\partial \nabla ^2,\partial \nabla ^{-1},\partial \nabla
^{-2}\right)\nonumber\\
{\cal R}^B&=&\unity \Box +{\cal R}_1\nonumber\\
{\cal R}_1&=&
\pmatrix{d_{\mu \nu \rho
}D^2\plg\partial &0&\kappa\partial  (D^{-2}u)\pl &0&0\cr
0&0&\kappa \partial D^4y_h & 0&0\cr
\kappa \partial\pl (D^{-2}u)\partial &0&0&0&\kappa \partial y_hD^{-2}\cr
0&0&0&\partial D^{-1}&0\cr}\ .
\end{eqnarray}
In the last expression $\Box$ is a flat $\partial \bar \partial $.

First for the calculation at $B=0$ we need for each entry the expression of
\begin{equation}
{\cal A}_j=-i\int dx\,dy\,\delta (x-y) (c\partial )_j G
(x,y;\ft{1}{M^2},\Phi^{(j)})\ ,
\end{equation}
with
\begin{equation}
\Phi ^{(j)}=\left\{ g^{\alpha \beta }=\pmatrix{-h &\ft{1}{2}\cr \ft{1}{2}&
0\cr},
{\cal Y}_+=0, {\cal Y}_- =-j(\partial h), E=-\ft12 j(\partial ^2h)\right\}\
.\label{Phij} \end{equation}
Using the Seeley--DeWitt coefficients\footnote{The
conventions are $ R^\alpha {}_{\beta \gamma \delta }=\partial _\gamma
\Gamma
^\alpha _{\beta \delta } -...$ and $R=R^\gamma {}_{\alpha \beta \gamma
}g^{\alpha \beta }$. Further $W_{\alpha \beta }=\partial _\alpha {\cal Y}
_\beta -\partial _\beta {\cal Y}_\alpha
+[{\cal Y}_\alpha ,{\cal Y}_\beta ]$.
} \begin{eqnarray}
\left.a_0\right|=1\ ; &\qquad & \left.\nabla _\alpha a_0(x,y)\right|=0
\nonumber\\
\left.a_1\right|= E-\ft16 R=\frac{1-3j}{6}(\partial ^2h)
\ ;&\qquad& \left.\nabla _\alpha
a_1(x,y)\right|=\ft12 \nabla _\alpha (E-\ft16 R) +\ft16 \nabla ^\beta
W_{\alpha \beta }
\ , \end{eqnarray}
where $|$ stands for the value at coincident points $x=y$ (after taking
the derivatives). The derivatives $\nabla _\alpha $ are covariant
w.r.t. gravitation.
We will need only $\nabla _+$ for which the
connection is zero:
\begin{equation}
\left.\partial a_1(x,y)\right|=\frac{1-3j}{12}(\partial
^3h)-\frac{j}{12}(\partial ^3h)\ . \end{equation}
This leads to
\begin{equation}
{\cal A}_j=\frac{-i}{24\pi }\int dx\, (6j^2-6j+1)c\,\partial ^3
h\ . \end{equation}
For the overall normalization of this anomaly, we used $\sqrt{g}=2$, in
accordance with the form of $g_{\alpha \beta }$ which follows from
\eqn{Phij}. In this way, we thus use
the coordinates $x^\alpha =\{  x^+,x^-\}$, and the
integration measure $dx$ in the above integral is then $dx^+\,dx^-$. If one
uses $dx=dx^0\,dx^1$ then the scalars as $R$ and $E$ do not change, but
$\sqrt{g}=\rho ^{-2}$, where $\rho $ is the parameter in the definitions in
footnote~\ref{fn:convrho}. Therefore the overall factor $1/(24\pi) $ in the
above formula, gets replaced by $1/(48\pi \rho ^2)$.
One can either interpret $dx$ as $dx^+\,dx^-$ or as $dx^0\, dx^1$ with
$\rho =1/\sqrt{2}$. The overall normalization in fact depends just on the
transformation law: if $S^1$ contains $X^*c\partial X$ then the anomaly for
one scalar is
\begin{equation}
{\cal A}_0=\frac{-i}{48\pi }\int dx^0\,dx^1\, c\,\partial (\partial
_0+\partial _1)^2 h\ , \end{equation}
independent of the definition of $\partial $. In all further expressions
for anomalies we again omit $\int dx$ with the normalization as explained
above.

Denoting the ghost combination in the transformation of the bosons as
\begin{equation}
\tilde c^\mu _\nu =c\delta ^\mu _\nu +2u\ds \plg\ , \label{deftildec}
\end{equation}
we obtain
\begin{equation}
\Delta ^hS = \Delta _{XX} ^hS+\Delta _{FF} ^hS=
\ft{1}{24\pi }(\tilde c^\mu _\mu-100\, c) \partial ^3h
\ ,\label{ANOMB0}
\end{equation}
where $\Delta _{XX}$ is the contribution from the matter entries in the
matrices, and $\Delta _{FF}$ comes from the fermions and gives the factor
$-100$.

For $\Delta ^BS$ we have to evaluate expressions as\footnote{The
calculation was first done without using this trick. Then the exponential
has to be evaluated using Campbell--Baker--Hausdorff expansions. The
manipulations described here simplify the calculations considerably and
were found in a discussion with Walter Troost.}
\begin{eqnarray} {\cal A}_B&=&tr\left[ K e^{t(\Box+{\cal R}_1)}\right] =
\frac{1}{t}\frac{d}{d\lambda } \left. tr\left[ e^{t(\Box+{\cal R}_1+\lambda
K)}\right] \right|_{\lambda =0}\nonumber\\
&=&
\frac{d}{d\lambda } \left. tr \left[ {\cal R}_1 e^{t(\Box+\lambda
K)}\right] \right|_{\lambda =0}\ ,
\end{eqnarray}
where the last step could be done because ${\cal R}_1$ is linear in $B$,
and we know that the result should be linear in $B$. We have for $K$ and
${\cal R}_1$ the general forms
\begin{equation}
K=k_0 +k_1\partial \ ;\qquad R=r_0+r_1\partial +r_2\partial ^2\ .
\end{equation}
Then we have to evaluate the heat kernel with
\begin{equation}
\Phi =\left\{ g^{\alpha \beta }=\pmatrix {0&\ft12\cr \ft12&0\cr},{\cal
Y}_+=0,{\cal Y}_-=\lambda k_1, E=\lambda (k_0-\ft12 \partial k_1)\right\}\
. \end{equation}
As the metric is flat, there is no non--trivial contribution from $\Delta
^{1/2}$ and $\sigma (x,y)$ in \eqn{Gexpbn}. The only new coefficient which
we need are the second derivatives at coincident points. For a flat metric,
and being interested only in linear terms in $E$ and ${\cal Y}$, the
coefficients are
\begin{equation}
\left.\nabla _\alpha \nabla _\beta a_0\right|=\ft12 W_{\alpha \beta }\
;\qquad
\left.\nabla _\alpha \nabla _\beta a_1\right|=\ft13 \partial
_\alpha \partial _\beta
E +\ft16 \partial ^\gamma \partial _{(\alpha }W_{\beta )\gamma }+{\cal O}
(\lambda ^2)\ , \end{equation}
where the symmetrization $(\alpha ,\beta )=\ft12 (\alpha \beta +\beta
\alpha )$. This gives
\begin{equation}
2\pi i{\cal A}_B= k_0\left(r_0-\ft12\partial r_1+\ft13\partial ^2r_2\right)
+(\partial k_1)\left(-\ft12r_0+\ft16\partial r_1-\ft1{12}\partial
^2r_2\right)\ , \end{equation}
which can be used to obtain
\begin{eqnarray}
\Delta ^BS&=&\Delta _{XX}^BS+\Delta _{XF}^BS+\Delta
_{FF}^BS\nonumber\\
i\Delta _{XX}^BS&=&
\frac{1}{12\pi }\tilde c^\mu _\nu
\partial ^3d_{\mu \nu \rho }B \plg \nonumber\\
i\Delta _{XF}^BS&=&-\frac{\kappa }{2\pi } (u\partial B-B\partial
u)(\partial ^3X^\mu )\pl \nonumber\\
i\Delta _{FF}^BS&=&\frac{\kappa }{6\pi }y_h\left\{ (-5+3\alpha
)u(\partial^3 B)+5(\partial ^3u)B\right.\nonumber\\
&&\hspace{1.6cm}\left. +(12-5\alpha )(\partial
u)(\partial ^2B) -12(\partial ^2u)(\partial B)\right\}\ .
 \label{ANOM}
\end{eqnarray}

We have thus obtained the anomaly at antifield number 0. It consists of
three parts. The first can be written in the form
\begin{equation}
i(\Delta S)^0_X= i\Delta _{XX}^hS+i\Delta _{XX}^BS=
\ft{1}{24\pi } \tilde c_{\mu \nu } \partial ^3\tilde h_{\mu \nu }\ ,
\label{cA0X}\end{equation}
where $\tilde c_{\mu \nu }$ was given in \eqn{deftildec}, and
\begin{equation}
\tilde h_{\mu \nu }=h\delta _{\mu \nu }+2d_{\mu \nu \rho }B\partial X^\rho
\ . \end{equation}
It is the total contribution from the matter loops, and agrees with
\cite{Hullmatter,SevrinQW3}. The other two parts originate in loops with
fermions. They are
\begin{eqnarray}
i(\Delta S)^0_F&=&i\Delta _{XF}S+i\Delta _{FF}^hS=
-\frac{100}{24\pi }\, c\, \partial ^3h
-\frac{\kappa }{2\pi } (u\partial B-B\partial
u)(\partial ^3X^\mu )\pl \nonumber\\
(\Delta S)^0_W&=& \Delta _{FF}^BS\approx 0\ .\label{cA0FW}
\end{eqnarray}
The upper index $0$ indicates that we have so far only the terms of
antifield number 0. These results agree with
\cite{SSvNMiami,Hullghost,Pope}.

\subsection{Consistency and antifield terms}
{}From the general arguments in section~\ref{ss:oneloop} and
appendix~\ref{app:consista} we know that the anomaly is consistent. At
antifield number zero this implies that $D^0 (\Delta S)^0\approx 0$. We may
check this now, and at the same time obtain the
anomaly at antifield number 1. This will e.g. include the contributions of
$h^*$, which according to \eqn{ctrW3gf} is the antighost. Note that this
subsection is just a check, and allows to compare with results of other
authors, but in principle the anomalies are determined and one may go
directly to section~\ref{ss:backgrch}.

It will turn out that the three parts mentioned above, $(\Delta S)^0_X$,
$(\Delta S)^0_F$ and $(\Delta S)^0_W$, are separately invariant under $D^0$
\begin{equation}
D^0(\Delta S)^0_X\approx 0\ \qquad
D^0(\Delta S)^0_F \approx 0\ ,   \label{consA}
\end{equation}
while this is obvious for $(\Delta S)^0_W\approx 0$.
To check this, one first obtains that
\begin{eqnarray}
D^0 \tilde c_{\mu \nu }&=&-\tilde c_{\rho (\mu }\partial \tilde
c_{\nu)\rho }+2\kappa
u(\partial u)\left[ 2\pl\plb +(\alpha +1)y_h\delta _{\mu \nu }\right]
\nonumber\\
D^0 \tilde h_{\mu \nu }&=&\left( \delta _{\rho(\mu  }\bar \partial
-\tilde h_{ \rho(\mu }\partial +(\partial \tilde h_{\rho (\mu })
\right)\tilde c_{\nu )\rho }\nonumber\\ &&
+2\kappa \left[2\pl\plb +\delta _{\mu \nu
}y_h\right](B\partial u-u\partial B)-2d_{\mu\nu \rho }y_\rho u\ .
\label{D0tilch}\end{eqnarray}
According to the theorem~\ref{thm:main}, this implies that the consistent
anomaly can be split in
\begin{equation}
\Delta S=(\Delta S)_X+(\Delta S)_F+(\Delta S)_W\ ,\label{DelSW3}
\end{equation}
where each term separately is invariant under ${\cal S}$, and starts with
the expressions in \eqn{cA0X} and \eqn{cA0FW}. Theorem~\ref{thm:main}
implies that the full expressions are obtainable from the consistency
requirement.

Indeed, from \eqn{consA} one can use \eqn{cSF00} to determine $
(\Delta S)_X^1, (\Delta S)_F^1$ and $(\Delta S)_W^1$: they are obtained by
replacing
the field equations $y_h$, $y_B$ and $y_X$ in the variation under $D^0$ by
$h^*$, $B^*$ and $X^*$. For the first one we obtain:
\begin{eqnarray}
i(\Delta S)_X^1&=&\frac{1}{24\pi }{\tilde c}_{\mu \nu }\partial
^3[-2\kappa
\delta _{\mu \nu }h^*(B\partial u-u\partial B)+2X_\rho ^*d_{\mu \nu \rho
}u]\nonumber\\&+&\frac{1}{24\pi }2\kappa u(\partial u)(1+\alpha )h^*\delta
_{\mu \nu }(\partial ^3{\tilde h}_{\mu \nu})\nonumber\\&+&\frac{1}{24\pi
}h^*[-8\kappa (\partial ^3c)(B\partial u-u\partial B)+8\kappa u(\partial
u)(\partial ^3h)] \label{AX1}
\end{eqnarray}
The first two terms can be absorbed in $(\Delta S)_X^0$ (\eqn{cA0X}) by
adding to ${\tilde c}_{\mu \nu }$ and ${\tilde h}_{\mu \nu }$:
\begin{eqnarray}
\tilde c^{(1)}_{\mu \nu }&=& 2\kappa(1+\alpha ) h^*
u(\partial u)\delta _{\mu \nu }
\nonumber\\
\tilde h^{(1)}_{\mu \nu }&=&-2\kappa \delta
_{\mu \nu }h^*(B\partial u-u\partial B)+2d_{\mu\nu \rho }X^*_\rho u\ ,
\end{eqnarray}
This part originated in the matter--matter entries of the
transformation matrix $K$ and the regulator $S_{\mu \nu }$ including all
antifields. If we consider these entries completely, we get as anomaly
\begin{equation}
i(\Delta S)_m= \ft1{24\pi } c_{\mu \nu } \partial ^3 h_{\mu \nu
}\label{cAm} \end{equation}
where
\begin{eqnarray}
c_{\mu \nu }&=&\tilde c_{\mu \nu }+\tilde c^{(1)}_{\mu \nu }\nonumber\\
h_{\mu \nu }&=&\tilde  h_{\mu \nu }+\tilde h^{(1)}_{\mu \nu }+2\kappa
(1-\alpha )c^*(\partial u)u\delta _{\mu \nu }\ . \end{eqnarray}
Note therefore that computing the matter anomaly by using just these
entries would not give rise to the last term of \eqn{AX1}:
\begin{equation}
i(\Delta S)_{X}-i(\Delta S)_m=\frac{\kappa }{3\pi
}h^*[- (\partial ^3c)(B\partial u-u\partial B)+ u(\partial
u)(\partial ^3h)] +\mbox{ terms of }afn\geq 2\ .
\end{equation}
As $\left(S,(\Delta S)_{X}\right) =0$, it follows that $(\Delta S)_m$
is not a consistent anomaly~! Indeed, the proof of consistency
given in appendix~\ref{app:consista} requires that we trace over all the
fields in the theory. One may check that the violation of the consistency
condition for $(\Delta S)_m$ agrees with \eqn{violSA}.
We will see that for $\alpha =0$ this extra term will be cancelled when
adding the fermion contributions.

For the other parts of the anomaly we obtain at antifield number~1:
 \begin{eqnarray}
\frac{6\pi i}{ \kappa}(\Delta S)_W^1&=&h^*\left\{\left(-49(\partial
^3c)+9(\partial ^2c)\partial \right)\left( u\partial B-B\partial u\right)
\right.\nonumber\\
&&\hspace{1cm}+\alpha \left[
u(\partial ^2c)\partial ^2B+10 (\partial u)B(\partial ^3c)+15
(\partial u)(\partial B)(\partial ^2c)\right.\nonumber\\&&\left.
\hspace{13mm} -15u(\partial ^3c)\partial B-6u(\partial ^4c)B\right]
\nonumber\\
&&\left.\hspace{1cm}+49(\partial ^3h)(\partial
u)u-9(\partial ^2h)(\partial ^2u)u+3\alpha
u(\partial ^3\nabla
u)-5\alpha (\partial u)(\partial ^2\nabla
u)\right\}\nonumber\\&&+
B^*\left\{ 9(\partial ^3u)(u\partial B-B\partial u)+(-9+\alpha )(\partial
u)u\partial ^3B+ 10\alpha (\partial ^2u)u\partial ^2B\right\} \nonumber\\
&&+X^*_\mu (\partial X^\mu )\left\{ 5u\partial ^3u+12(\partial ^2u)\partial
u\right\} \nonumber\\[3mm]
\frac{6\pi i}{ \kappa}(\Delta S)^1_F&=&-50 h^*\left\{ (1-\alpha )(\partial
u)u\partial ^3h+(\partial ^3c)(B\partial u-u\partial B)\right\} \nonumber\\
&& -3h^*\left\{  \left( (\partial ^3c)+3 (\partial ^2c)\partial \right)
(u\partial B-B\partial u)
+u(\partial u)\partial ^3h+3u(\partial ^2u)\partial
^2h\right\} \nonumber\\
&&-3B^*\left\{ -3(u\partial B-B\partial u)\partial
^3u+3u(\partial u)\partial ^3B\right\} \nonumber\\
&&-3X_\mu ^*\left\{ -2u(\partial u)\partial ^3X^\mu -(\partial
X^\mu )(\partial (u\partial ^2u))-2u(\partial ^2u)\partial ^2X^\mu \right\}
\end{eqnarray}
Remarkable simplifications occur for the full anomaly:
\begin{eqnarray}
i(\Delta S)^0+i(\Delta S)^1&=& i(\Delta S)_m^0+i(\Delta S)_m^1+i(\Delta
S)_F^0+ i(\Delta S)_W^0\nonumber\\
&&+\frac{\kappa }{6\pi }X^*_\mu \left\{ 6\partial (u(\partial u)\partial
^2X^\mu)+9(\partial ^2u)(\partial u)\partial X^\mu +8u(\partial
^3u)\partial X^\mu \right\}  \nonumber\\
&&
+\frac{\alpha \kappa }{6\pi }h^*\left\{
u(\partial ^2c)\partial ^2B+10 (\partial u)B(\partial ^3c)+15
(\partial u)(\partial B)(\partial ^2c)\right.\nonumber\\&& \hspace{13mm}
 -15u(\partial ^3c)\partial B-6u(\partial ^4c)B\nonumber\\
&&\left.\hspace{13mm}+50(\partial u)u\partial ^3h+3u\partial ^3\nabla
u+5(\partial ^2\nabla
u)\partial u\right\} \nonumber\\&&+\frac{\alpha \kappa }{6\pi
}B^*\left\{ (\partial u)u\partial ^3B+10(\partial ^2u)u\partial
^2B\right\} \ , \end{eqnarray}
where $(\Delta S)_m^0$ and $(\Delta S)_m^1$ are the terms of antifield number 0
and 1 in \eqn{cAm}. Note especially the simplification when using the
parametrization with $\alpha =0$. The terms with the `antighost' $h^*$ are
included in the `matter anomaly' $(\Delta S)_m$, \eqn{cAm}, and $B^*$
disappears completely.
\vspace{1cm}

Let us recapitulate what we have determined. First remark that the
regularization depends on an arbitrary matrix $T$, and this implies that,
not specifying $T$,  $\Delta S$ is only determined up to $(G,S)$, where $G$
is a local integral \cite{anombv}. We have chosen
a regularization (a specific matrix $T$). This determines the value of
$\Delta S$. However, we have calculated only the part of $\Delta S$ at
antifield number 0 (including the weakly vanishing terms). If we would have
calculated up to field equations (which would in principle be sufficient
to establish whether the theory has anomalies), then we would have
determined
$\Delta S$ up to $(G,S)$, where $G$ has only terms with antifield
number~1 or higher.  In our calculation of section~\ref{ss:calc1lW3}, we
determined also the weakly vanishing terms. Therefore the value of $\Delta
S$ has been fixed up to the above arbitrariness with terms $G$ of antifield
number~2 or higher. Indeed, looking at \eqn{cSF00} one can
always shift $(\Delta S)^1\rightarrow (\Delta S)^1+\delta _{KT}G^2$, for
some arbitrary function $G^2$ of antifield number two. This arbitrariness
can also be seen from the fact that changing $T$ at a certain
antifield number, would change the counterterms starting at the same
antifield number. Our calculation above did not depend on possible new
terms
of $T$ at antifield number 2 or higher (changing $T$ at antifield number 1,
would change the weakly vanishing terms of the anomaly). Of course,
the remaining arbitrariness is not important when we want to investigate
whether a theory has anomalies.

To obtain the complete form of $\Delta S$ up to $(S,G)$ one can continue
the calculations of this subsection to determine the terms of antifield
number 2.

\section{Background charges in $W_3$ gravity}
\label{ss:backgrch}
\setcounter{equation}{0}
It is by now well--known that the anomalies can be cancelled in chiral
$W_3$ gravity by including background charges \cite{ROMANS}. We will see in
this section how this can be implemented in the BV language.

To cancel the anomalies by local counterterms, we first note that $(\Delta
S)^0_W\approx 0$, and theorem~\ref{thm:main} thus implies that there is a
local counterterm, which starts with (we take in this section $\alpha =0$,
but of course
these steps can also be done in parametrizations with $\alpha \neq 0$, as
this involves only a canonical transformation). \begin{equation}
M_{W1}{}^1=-\frac{\kappa }{\pi } B\left[ \frac{5}{6}u(\partial
^3h^*)+\frac{9}{2}(\partial u)(\partial ^2h^*)+\frac{17}{2}(\partial
^2u)(\partial h^*)+\frac{17}{3}(\partial ^3u)h^*\right]\ , \label{counter1}
\end{equation}
which is the right hand side of the last expression in \eqn{ANOM}, with
$y_h$ replaced by $h^*$, where we added a total derivative for later
convenience.
The other terms in \eqn{DelSW3} can not be countered by a local integral.

Background charges are terms with $\sqrt{\hbar }$ in the (extended) action.
We thus have
\begin{equation}
W=S+\sqrt{\hbar}\,M_{1/2}+\hbar M_1+\ldots \ .
\end{equation}
Therefore the expansion \eqn{MEC} of the master equation \eqn{ME} is now
changed to
\begin{eqnarray}
(S,M_{1/2})&=&0 \label{Mhalf}\\
(S,M_1)&=&i\Delta S-\frac{1}{2}(M_{1/2},M_{1/2})\,. \label{MEsqh}
\end{eqnarray}
Relevant terms $M_{1/2}$ are those which are in the antibracket
cohomology. Indeed, if $M_{1/2}$ which solves \eqn{MEsqh} has a part
$(S,G)$, then
we find also a solution by omitting that part of $M_{1/2}$, and adding to
$M_1$ a term $\ft12\left( M_{1/2},G\right) $.
These terms $M_{1/2}$ are thus again determined by their part at antifield
number zero. In chiral $W_3$ gravity one considers
\begin{equation} M_{1/2}{}^0=(2\pi )^{-1/2}\left[ a_\mu
h(\partial ^2X^\mu )+e_{\mu \nu }B\pl (\partial ^2 X^\nu )\right] \ ,
\end{equation}
where the numbers $a_\mu $ and $e_{\mu \nu }$ are the background charges,
and the numerical factor is for normalization in accordance with previous
literature. We first consider \eqn{Mhalf}.
Using our theorem, $D^0M_{1/2}{}^0$ should be weakly zero
in order to find a solution. This gives the
following conditions on the background charges:
\begin{eqnarray}
e_{(\mu \nu )}-\ds a_\rho&=&0\nonumber\\
\ds (e_{\rho \sigma }-e_{\sigma \rho })+2e_{(\mu }{}^\rho d_{\nu )
\sigma \rho }&=& b_\sigma \delta _{\mu \nu }\ , \label{condbM12}
\end{eqnarray}
where $b_\sigma $ is determined to be $2\kappa a_\sigma $ by the
consistency of the symmetric part
$(\mu \nu \sigma )$ of the last equation with the first one and
\eqn{dsym}. If these conditions are satisfied,
we know that we can construct
the complete $M_{1/2}$ perturbatively in antifield number. The solution is
\begin{eqnarray}
M_{1/2}&=&
M_{1/2}{}^0+M_{1/2}{}^1+M_{1/2}{}^2
\nonumber\\
M_{1/2}{}^1&=&(2\pi )^{-1/2}\left[  -a_\mu X^*_\mu (\partial
c)+e_{\mu
\nu }X^*_\mu u(\partial ^2X^\nu )+e_{\nu \mu }(\partial X^*_\mu) u(\partial
X^\nu)\right.\nonumber\\
& &\hspace{25mm}\left.+\kappa a_\mu h^*(D^{-2}u)(\partial ^2X^\mu
)\right] \nonumber\\
M_{1/2}{}^2&=&(2\pi )^{-1/2}2\kappa a_\mu\left[ (\partial
X^*_\mu) h^* - c^*(\partial ^2X^\mu )\right]u (\partial u)\ .
\end{eqnarray}
We calculated here the terms of antifield number 2 (and checked that there
are no higher ones), but note that these are not necessary for the analysis
below.

In the set of solutions of \eqn{dsym} as explained in
footnote~\ref{fn:soldsym}, we have shown that there is only a solution for
these equations in the case $D=1$ and $r$ arbitary \footnote{In
\cite{ROMANS} this solution was found, and no solution was found for the
other cases, but this was also not excluded. Now we have confirmed that the
other cases, with the 4 `magical solutions', do not allow a solution.}.
There is thus exactly one solution for each value of $n$, the range of the
index $\mu $. For these models, the solution of \eqn{dsym} can be simply
written as
\begin{equation}
d_{111}=-\sqrt{\kappa }\ ;\qquad d_{1ij}=\sqrt{\kappa }\,\delta _{ij}\ ,
\label{solD1} \end{equation}
where $i=2, \ldots, n$. The solution to \eqn{condbM12} is
\begin{eqnarray}
&&e_{00}=-\sqrt{\kappa }\,a_0\ ;\qquad
e_{ij}=\sqrt{\kappa }\,a_0\delta _{ij}
\nonumber\\
&&e_{0i}=2\sqrt{\kappa }\,a_i\ ;\qquad  e_{i0}=0\ , \label{solD1e}
\end{eqnarray}
where $a_\mu $ is arbitrary.

The final relation for absence of anomalies up to one loop is that we have
to find an $M_{1}$ such that the last equation of \eqn{MEsqh} is satisfied.
Again we only have to verify this at zero antifield number, and up to field
equations of $S^0$, due to theorem~\ref{thm:main}. We calculate
$Q\equiv i\Delta
S-\frac{1}{2}(M_{1/2},M_{1/2})$ at antifield number zero. It contains
terms proportional to $ c\partial ^3h$, which can not be removed by a
local counterterm. So in order to have no anomaly, the
multiplicative factor of this term has to vanish.
This implies the relation
\begin{equation}
c_{mat}\equiv n-12a_\mu a_\mu =100
\end{equation}
For the other terms in $Q$, one needs a counterterm
\begin{equation}
M_{B1}{}^0=\frac{1}{6\pi }e_{\mu \nu }a_\mu B\partial ^3 X\ ,
\end{equation}
and one imposes the relations
\begin{eqnarray}
&&2e_{\mu \nu }a_\mu -6a_\mu e_{\nu \mu }+d_{\nu \mu \mu }=0\nonumber\\
&&e_{\mu \rho }e_{\rho \nu }=\kappa Y\delta _{\mu \nu }\nonumber\\
&&-e_{\mu \rho }e_{\nu \rho }+\ft13 d_{\mu \rho \sigma }d_{\nu \rho \sigma
}+\ft23d_{\mu \nu \rho }e_{\sigma \rho }a_\sigma =\kappa \delta _{\mu \nu
}Z\nonumber\\
&&-4 \kappa a_\mu a_\nu +e_{\rho \mu }e_{\rho \nu }+2a_\rho e_{\rho \sigma
}d_{\sigma \mu \nu }=(3Z+4Y-2)\kappa \delta _{\mu \nu }\ , \end{eqnarray}
where $Y$ and $Z$ are arbitrary numbers, to be determined by consistency
requirements.
This set of equations on the background charges are exactly
the same as in \cite{ROMANS}.
The solutions \eqn{solD1} and \eqn{solD1e} give now
\begin{eqnarray}
&&a_0{}^2=-\frac{49}{8}\ ;\qquad  a_i{}^2= \frac{-53+2n}{24}\nonumber\\
&&Y=a_0{}^2\ ;\qquad Z=\frac{1}3 \left(2-a_0{}^2\right)=\frac{87}{8}\ .
\end{eqnarray}

Then we obtain that
\begin{eqnarray}
&&2\pi \left( (M^0_{1/2},M^1_{1/2})-i\Delta
S+(M_{B1}^0,S^1)\right) =-\ft13 e_{\nu \mu }a_\nu (\partial
^2u)y_\mu\\
&&+\kappa B\left( 2Z (\partial ^3u) y_h +3Z(\partial ^2u)\partial
y_h+ (3Z-2
+2Y)(\partial u)\partial ^2y_h+(Z+Y-1 )u\partial ^3y_h\right) \ .
\nonumber \end{eqnarray}
This determines then the counterterm $M_{B1}$ at antifield number 1.
Inserting the values for $Y$ and $Z$ gives
\begin{eqnarray}
M_{B1}{}^1&=&\frac{\kappa }{16\pi } B\left[ 30u(\partial
^3h^*)+147(\partial u)(\partial ^2h^*)+261(\partial
^2u)(\partial h^*)+174(\partial ^3u)h^*\right]\nonumber\\&&+
\frac{1}{6\pi }X^*_\mu e_{\nu \mu }a_\nu \partial ^2u\ .
\label{counter2} \end{eqnarray}
Combining this term with \eqn{counter1}, we get for the total
counterterm at antifield number one
\begin{eqnarray}
M_1{}^1&=&
\frac{25\kappa }{48\pi } B\left[ 2u(\partial
^3h^*)+9(\partial u)(\partial ^2h^*)+15(\partial
^2u)(\partial h^*)+10(\partial ^3u)h^*\right] \nonumber\\
&&+\frac{1}{6\pi }X^*_\mu e_{\nu \mu }a_\nu \partial ^2u\ .
\label{counter}
\end{eqnarray}
The form of this counterterm is also obtained in
\cite{Pope,SSvNMiami,Bergshoeff}. As mentioned before, this does not only
determine the quantum corrections to the action, but also the corrections
to the BRST transformations, by looking to the linear terms in antifields
in the gauge--fixed basis.

The value of $\kappa $ has been irrelevant here. In fact, one can remove
$\kappa $ rescaling $d _{\mu \nu \rho }$, $e_{\mu \nu }$, $B^*$ and
$u^*$ with $\sqrt{\kappa }$ and $B$ and $u$ by $(1/\sqrt{\kappa })$. For
the usual normalizations in operator product expansions of the $W$--algebra
one takes
\begin{equation}
\kappa =\frac{1}{6Z}=\frac{8}{22+5c_{mat}}=\frac{4}{261}\ .
\end{equation}
\section{Conclusions}
By applying the BV formalism and the regularization procedure initiated in
\cite{anomPV,measure,anombv} to chiral $W_3$ gravity, we have shown how it
can be applied to more difficult theories. Gauge fixing could be performed
by the most simple canonical transformation from the classical extended
action. We have
seen how the theorem on antibracket cohomology can be effectively used in
calculations.
Background charges have been introduced as terms of order
$\sqrt{\hbar }$. They lead to a new expansion of the quantum master
equation: \eqn{Mhalf}, \eqn{MEsqh}. This method allows an easy verification
of the cancellation of anomalies.

We performed only the one--loop calculations. The higher loops should be
regulated using higher derivative regularization (still combined with
Pauli--Villars) \cite{FadSlav}. This is again a regularization within the
path integral. Recently it
has been successfully used in Wess--Zumino--Witten models to obtain the
anomalies at all loops \cite{ARW}.

It would be interesting to consider also $W_3$ gravity as an example for
higher loop anomalies. Usually in gauge theories one may use
covariant derivatives to built the terms with higher derivatives for the
regularization. Then these terms are invariant under the gauge
transformations, and therefore they produce no anomalies. The only
anomalies which occur, originate in the Pauli--Villars mass terms, which
are
genuine one--loop anomalies. This situation is different in $W$ gravity,
because in these theories there are no covariant derivatives (with a finite
number of terms). Therefore one expects here new anomalies from higher
loops, which indeed are known to appear \cite{Matsuo}.

\vspace{1cm}

\section*{Acknowledgments.}
We thank Walter Troost for several very helpful discussions.
\newpage

\section*{Appendix}
\appendix
\section{Consistency of the regulated anomaly}
\label{app:consista}
\setcounter{equation}{0}
The PV regularization should give a consistent anomaly, as all
manipulations can be done at the level of the path integral. Here we want
to check this explicitly from the final expression for the anomaly after
integrating out the PV fields. Then we will consider the expression which
is obtained after integrating out only parts of the fields. We will see
that in that case the resulting expression does not satisfy this
consistency equation.

The expression for the anomaly depends on an invertible matrix $T_{AB}$:
\begin{equation}
\Delta S = str\left[ J \frac{1}{\unity - {\cal R}/M^2}\right]\
,\label{DelSreg} \end{equation}
where
\begin{eqnarray}
K^A{}_{B}=\dl^A  S\dr_B\ ;&\qquad& \underline{S}_{AB}=\dl_A S\dr_B
\nonumber\\
J=K+\ft12 T^{-1}({\cal S}T)\ ;&\qquad&{\cal R}=T^{-1}\underline{S}\ ,
\label{defKSJO}\end{eqnarray}
and for matrices $M^A{}_B$ or $M_{AB}$ we define the nilpotent operation
\begin{equation}
\left( {\cal S}M\right) _{AB}= \left( M_{AB},S\right) (-)^B\ .
\end{equation}
Matrices can be of bosonic or fermionic type. The Grassmann parity
$(-)^M$ of a matrix is the statistic of $M^A{}_B(-)^{A+B}$. So of the above
matrices, $J$, $K$ and $({\cal S}T)$ are fermionic, the other are bosonic.
The definition of supertraces and supertransposes depend on the position of
the indices.
\begin{eqnarray}
\left( R^T\right) _{BA}=(-)^{A+B+AB+R(A+B)}R_{AB}\ ;&\qquad &
\left( T^T\right) ^{BA}=(-)^{AB+T(A+B)}T_{AB} \nonumber\\
\left( K^T\right)_B{}^A=(-)^{B(A+1)+K(A+B)}K^A{}_{B}\ ;&\qquad &
\left( L^T\right)^A{}_B=(-)^{B(A+1)+L(A+B)}L_B{}^A\nonumber\\
str\ K = (-)^{A(K+1)} K^A{}_{A}\ ;&\qquad&
str\ L = (-)^{A(L+1)} L_A{}^A
\ . \label{Trules}
\end{eqnarray}
This leads to the rules
\begin{eqnarray}
&&\left( M^T\right) ^T =M\ ;\qquad (MN)^T =(-)^{MN} N^T M^T\ ;\qquad
\left( M^T\right) ^{-1}=(-)^M\left( M^{-1}\right) ^T\nonumber\\
&&str\ (MN)= (-)^{MN}str\ (NM)\ ;\qquad str\ (M^T)= str\ (M)\nonumber\\
&&{\cal S}(M\,N)=M({\cal S}N)+(-)^N ({\cal S}M)N\ ;\qquad
{\cal S}\,(str\,M)=str\ ({\cal S}M)\,.
\label{propstr}\end{eqnarray}
The second derivatives of the master equation lead to
\begin{eqnarray}
{\cal S}\underline{S} &=& -K^T \underline{S}-\underline{S}\,K\nonumber\\
{\cal S}K&=&  \overline{S}\,\underline{S}-K\,K \ ,
\end{eqnarray}
where,
\begin{equation}
\overline{S}=\dl^A  S\dr^B\ ;\qquad
(K^T)_A{}^B=-\dl_A  S\dr^B\ ,
\end{equation}
the first being a graded antisymmetric matrix, and the second is in
accordence with \eqn{defKSJO} and the previous rules of supertransposes.
We rewrite the expression of the anomaly as
\begin{equation}
\Delta S=M^2\ str\left[ T\,J\, P^{-1}\right] \ ,
\end{equation}
where
\begin{equation}
P=M^2\,T-\underline{S}\ .
\end{equation}
We also easily derive the following properties
\begin{eqnarray}
{\cal S}(T\,J)&=&{\cal S}(T\,K)=T\overline{S}\underline{S}-TKK-({\cal
S}T)K\nonumber\\
{\cal S}P&=&M^2(TJ+J^TT)-(PK+K^TP) \label{varyP}\ .
\end{eqnarray}
This leads to
\begin{eqnarray}
{\cal S}\Delta S&=&M^2\
str\left[\left(T\overline{S}\underline{S}-TKK-({\cal
S}T)K\right)P^{-1}\right.\nonumber\\&&\hspace{16mm}\left.-TJP^{-1}\left(M^2
( T
J + J ^ T T ) - ( P K + K^T P ) \right) P ^ { - 1 } \right] \end{eqnarray}
In the first term of the first line we write $\underline{S}=M^2T-P$. The
trace of both these terms is zero due to \eqn{propstr} and the
(a)-symmetry properties of the matrices given above. For the same reason
the second term of the second
line vanishes. The first term of
the second line is a square of a fermionic matrix which vanishes under the
trace. The remaining terms again combine into matrices which are traceless
by using their symmetry and by \eqn{propstr}.
This means we have proven that PV-regularization guarantees
consistent one--loop anomalies,
\begin{equation}
{\cal S}\Delta S=0\ .
\end{equation}
In \cite{JGJPII} a formula has been given for the non--local counterterm
for any $\Delta S$ defined by \eqn{DelSreg}:
\begin{equation}
\Delta S=-\frac{1}{2}{\cal S}\left( str \ln \frac{{\cal R}}{M^2-{\cal R}
}\right) \ . \end{equation}
This can be proven also from the above formulas.
\vspace{1cm}

Consider the part of the anomaly originating from the path integral over
some subset of fields, e.g. the matter fields in $W_3$.
The question arises whether this gives already a consistent anomaly.
To regulate this anomaly we only have to introduce
PV--partners for this subset of fields. In the space of all
fields, and taking the basis with first the fields which are integrated, we
write the $T$--matrix as \begin{eqnarray}
T=\left(\begin{array}{cc}
\tilde T & 0\\
0 & 0
\end{array}\right)\ .
\end{eqnarray}
Because only this subsector is integrated, we have to project out the
other sectors (mixed and external) in the full matrix of second
derivatives $\underline{S}$ before inverting it to define a
propagator. This can be done by defining the projection operator $\Pi $ as
\begin{eqnarray}
\Pi =\left(\begin{array}{cc}
\unity & 0\\
0 & 0
\end{array}\right)\ .
\end{eqnarray}
The inverse propagator is then $Z=(\Pi P\Pi)$. Further we understand by
inverses, as e.g. $T^{-1}$, the inverse in the subspace. So we
have \begin{equation}
TT^{-1}=T^{-1}T=\Pi \ .
\end{equation}
The matter anomaly takes the form
\begin{equation}
(\Delta S)_m=M^2\ str[TJZ^{-1}]\ ,
\end{equation}
and \eqn{varyP} remains valid. In the variation of this anomaly, we now
often encounter
\begin{equation}
P\,Z^{-1}=\Pi + Y \qquad \mbox{where}\qquad Y=-(1-\Pi )\underline{S}Z^{-1}\
. \end{equation}
Using again the properties of supertraces and transposes, we obtain
\begin{equation}
{\cal S}(\Delta S)_m=M^2\ str\left[-T\overline{S}Y-TK(\unity -\Pi
)KZ^{-1}-Y^TKZ^{-1}(TJ+J^TT)\right] \ .\label{violSA}
\end{equation}
This thus shows that one does in general not obtain a consistent anomaly
when integrating in the path integral only over part of the fields.  One
can see that for ordinary chiral gravity the structure of the extended
action
implies that each term of \eqn{violSA} vanishes.  For chiral $W_3$ gravity
however, some terms remain.  A similar result for Ward identities was
obtained in \cite{FBWardid}.

\section{Lemmas and proofs concerning the KT acyclicity.}
\label{app:lemprKT}
\renewcommand{\thelemma}{\thesection.\arabic{lemma}}
\renewcommand{\thetheorem}{\thesection.\arabic{theorem}}
\setcounter{equation}{0}
\setcounter{lemma}{0}
\setcounter{theorem}{0}
\begin{lemma} :
After a canonical transformation, the
new KT differential is expressed in terms of the old one as follows
\begin{equation}
\dkt' F(\Phi'{}^*,\Phi ') =\left.\dkt \Phi '{}^*_A\right|_{c'=0} \dl{}'{}^A F
\ .\label{dktpdkt}
\end{equation}
The primed derivatives are taken with all other primed fields
constant, and also the notation $c'=0$ means that the considered
expression is first written in terms of primed variables, and then all
`ghosts' are put equal to zero. The word `ghosts' stands for all
fields with positive definite ghost number. \end{lemma}
First, we can use the following definition of the KT differential
\begin{equation}
\dkt F(\Phi ^*,\Phi )= \left(S\dr_A\right)_{c=0} \dl{}^A F\ .
\label{defdkt2}
\end{equation}
One may check that this is equivalent with \eqn{defdkt0}. Now we
obtain for the primed KT differential
\begin{eqnarray}
\dkt' F(\Phi'{}^*,\Phi' )&=& \left(  S\dr'_A\right) _{c'=0} \dl{}'{}^A
F\nonumber\\
&=&\left(  S\dr_B\cdot \left. \Phi ^B\dr'_A\right|_{\Phi '{}^*}
+ S\dr{}^B\cdot \left. \Phi^*_B\dr'_A\right|_{\Phi '}
\right) _{c'=0} \dl{}'{}^A F\ .
\end{eqnarray}
In the last expression the second term in the brackets can be omitted
because $\partial ^B S$ has positive ghost number, and will thus
vanish when all fields of positive ghost number in the primed basis
are put equal to zero. (Canonical transformations conserve the ghost
number). For the first term we use eq.(A.30) in \cite{anombv}
\begin{equation}
\left. \Phi ^B\dr{}'_A\right|_{\Phi '{}^*} = \left( M^{-1}\right)
^B{}_{A}= \left. \dl{}^B \Phi '{}^*_A\right|_{\Phi }\ . \end{equation}
Therefore this term is by definition equal to $\dkt \Phi '{}^*_A$, which
gives the result \eqn{dktpdkt}\QED

The previous lemma implies that for any function $F(z')$
\begin{equation}
\dkt ' F(z') = \left(\dkt F(z'(z))\right) _{c'=0}+\
c'\mbox{--dependent terms}\ .\label{lemnewform} \end{equation}
The only non--trivial step in this new formulation of the lemma is that
$\dkt \Phi '{}^A$ gives only $c'$--dependent terms.

\begin{theorem} :
If the KT differential is acyclic in one set of coordinates, then it
is also acyclic after a canonical transformation.
\label{lemKTctr}\end{theorem}
Suppose the KT differential is acyclic in primed coordinates. Then
consider $F(z_0)$ such that $\dkt F(z_0)=0$, where we use the notation
$z=\{\Phi ^*,\phi ,c\}$ and $z_0$ is the same set with $c=0$.
Now define \begin{equation}
F'(z '_0)\equiv F(z_0(z'_0))
\end{equation}
where $z(z')$ is the canonical transformation. This function is thus
just rewriting $F$ in primed coordinates and then putting $c'=0$, thus
\begin{equation}
F(z_0(z'))=F'(z'_0)+c'\mbox{--dependent terms.} \label{defFFp}
\end{equation}
 From our starting point we thus have
\begin{equation}
0=\dkt F'(z'_0)+\dkt c'\mbox{--dependent terms}\ ,    \label{0dktFp}
\end{equation}
and the last term is still proportional to $c'$ because of ghost
number conservation of the canonical transformation, and $\dkt c =0$.

Now we use \eqn{lemnewform} for $F'$.
As $F'$ does not depend on $c'$, also $\dkt' F'$ does not depend on
$c'$, and therefore the first term gives the complete result.
Eq. \eqn{0dktFp}
with the remark following it, then implies that $\dkt' F'=0$.

The assumption of acyclicity of $\dkt'$ implies that
\begin{equation}
F'(z'_0)=\dkt ' H'(z'_0) + h'(\phi ')\ .
\end{equation}
We now follow the same method as above with primed and unprimed fields
interchanged. We define
\begin{eqnarray}
H(z_0)&\equiv& H'(z'_0(z_0))\nonumber\\
H'(z'_0(z))&=& H(z_0)+\ c\mbox{--dependent terms }\nonumber\\
\dkt' H'(z'_0(z))&=&\dkt' H(z_0)+\ c\mbox{--dependent terms .}
\end{eqnarray}
Now we use the previous lemma in the reverse order
\begin{eqnarray}
\dkt H(z_0) &=& \left( \dkt' H(z_0)\right) _{c=0}\nonumber\\
&=& \left( \dkt' H'(z'_0(z)\right) _{c=0}\nonumber\\
&=& \left( F'(z'_0)-h'(\phi ')\right) _{c=0}
\end{eqnarray}
and then with \eqn{defFFp} this implies (at $c=0$, the transformation
$\phi '(z)$ is restricted to $\phi '(\phi )$) \begin{equation}
F(z_0)=\dkt H(z_0) + h'(\phi '(\phi ))\ .
\end{equation}
If the
definition of the functions ${\cal F}_s^0$ has been taken in both
variables such that the subpart of the
canonical transformation (the fields at ghost number zero) $\phi
'(\phi )$ leaves this set of functions invariant, this finishes the
proof. Otherwise, one uses lemma~\ref{lem:dkt0r} to write the difference
as another $\dkt$--exact expression. This proves the acyclicity in
unprimed coordinates. \QED

\begin{lemma} :
One can perform a canonical transformation such that $f^{ij}{}_{a_1}$
defined by \eqn{deffa1} does not have parts proportional to
$R^i{}_{a}$ or $R^j{}_{a}$.\label{lemctrfij}\end{lemma}
The exact definition of `proportional' is explained in
lemma~\ref{lem:ac2}. The extended action is
\begin{eqnarray}
S&=& S^0(\phi )+\phi ^*_i R^i{}_{a}c^a \nonumber\\ &&+\left( c^*_a
Z^a{}_{a_1}+(-)^i\phi ^*_i\phi ^*_jf^{ji}{}_{a_1}\right)
c^{a_1}+\ldots \label{Supto2} \end{eqnarray}
(where $\ldots$ are terms of antifield number 3 or more, or quadratic in
ghosts). We consider now the case that
\begin{equation}
(-)^i\phi ^*_i \phi ^*_j
f^{ji}{}_{a_1}=(-)^ay_ay_bg^{ba}{}_{a_1}+(-)^i\phi
^*_iy_a g^{ai}{}_{a_1} +(-)^i\phi ^*_i \phi ^*_j g^{ji}{}_{a_1}\ .
\end{equation}
where $y_a=\phi ^*_i R^i{}_{a}$. We perform
a canonical transformation  using the generating function (the
transformation being defined by \eqn{cantrf}) \begin{equation}
f=(-)^a c'{}^*_a  \phi '{}^*_i\left(
g^{ja}{}_{a_1}+R^j{}_{b}g^{ba}{}_{a_1}\right) c^{a_1}\ .
\end{equation}
One may check that the
action in terms of the primed fields is again as in \eqn{Supto2}, with
$f$ replaced by $g$.
\QED

First we will prove acyclicity in ${\cal F}^*_1$:
\begin{lemma}: The KT differential is acyclic on functions
of $\phi ^i$ and $\phi ^*_i$, modulo terms
proportional to $\phi ^*_iR^i{}_a$.\label{lem:dkt0r}
\end{lemma}
First define an operator $d$ which is
\begin{equation}
dF = \phi ^*_i \frac{\dl}{\partial y_i} F + \phi ^*_i R^i{}_a \epsilon ^a\
. \end{equation}
The derivative w.r.t. $y_i$ was defined in \eqn{dFdy}, where we noticed
that
its value is undetermined, corresponding to an arbitrary function $\epsilon
^a$ in the above equation. In ${\cal F}^*_1$, we have
\begin{equation}
\dkt F  =y_i\frac{\dl}{\partial \phi ^*_i}F\ .
\end{equation}
This implies that
\begin{equation}
\left( \dkt d+d\dkt\right)F = y_i\frac{\dl }{\partial y_i}F +\phi ^*_i
\frac{\partial }{\partial \phi ^*_i} F + \phi ^*_i R^i{}_a \epsilon '{}^a \ .
\end{equation}
Now we expand any function as well according to \eqn{Fexpy} as in its
number of antifields. This defines
\begin{equation}
F=\sum_{m,n=0}^{\infty } F^{m,n}\ ,\end{equation}
where $m$ counts the number of antifields and $n$ the number of field
equations in each term.
We have now also
\begin{equation}
\left( \dkt d+d\dkt\right)F^{m,n} =(m+n) F^{m,n}+ \phi ^*_i R^i{}_a \epsilon
'{}^a.
\end{equation}
As $\dkt$ does not change the value of $m+n$, any function which
vanishes under $\dkt$, also has $\dkt F_N=0$, where $F_N$ has all
terms with $m+n=N$. Therefore with the previous formula, $F_N$ for
$N>0$ is the
$\dkt $ of something, modulo $\phi ^*_i R^i{}_a \epsilon ^a$. The functions
with $m+n=0$ are those of ${\cal F}^0_s$. We thus obtain for $F\in {\cal
F}^*_1$ \begin{equation}
\dkt F=0 \ \Rightarrow\ F=\dkt H +\phi ^*_i R^i{}_a \epsilon ^a+f(\phi )\
\mbox{ with } f(\phi )\in {\cal F}^0_s\ .
\end{equation}
\QED
Now we will go one step further. The relations \eqn{RZisf} with
\eqn{deffa1} will be important.
\begin{lemma} :
The KT differential is also acyclic on ${\cal F}^*_{2}$ (antifields
up to $c^*_{a}$), modulo terms
proportional to
\begin{equation}
\dkt c^*_{a_1}=c^*_{a}Z^a{}_{a_1}+(-)^i\phi ^*_i\phi
^*_jf^{ji}{}_{a_1}\ . \end{equation}
\label{lem:ac2}\end{lemma}
We define $y_a\equiv \phi ^*_i R^i{}_{a}$. We will expand functions
$F(\phi^i ,\phi ^*_i, c^*_a)$ in powers of $c^*_a$ and $y_a$.
The first expansion is obvious. The expansion in $y_a$ is similar to
the one in $y_i$ explained before. We have to choose a representant
between the antifields $\phi ^*_i$ to define when we have terms
proportional $y_a$. E.g. for the 2--dimensional gravity model of
section~\ref{ss:ingrBV}, the second term in \eqn{SextW2} is $yc$ (where
here $y\equiv y_a$). We choose here to look first at the terms
with $h^*$ and write all terms proportional to $\bar
\partial h^*$ as \begin{equation}
\bar \partial h^*=-y+X ^*_\mu \, \partial X^\mu  +
\partial (h^*h) +h^*(\partial h)\ .
\end{equation}

The relation $y_i(F^i-G^i)=0$ defines again an equivalence relation
between functions $F^i(\phi )$ with a free index of type $i$. The
properness condition \eqn{allsym} gives the possibilities for the
difference $F^i-G^i$. We further reduce these functions by identifying
those which are equal on the stationary surface for all $i$. The above
procedure
defines for each equivalence class one representative function.
We denote the set of these representative functions as ${\cal
F}^1_{s}$. We thus obtain
\begin{equation}
y_i(F^i-G^i)=0\ \mbox{ or }F^i\approx G^i\mbox{ where } F^i,G^i \in
{\cal F}^1_{s}\ \Longrightarrow\ F^i=G^i\ .
\end{equation}
We can also generalize this to functions with more free indices of
type $i$. Then the above requirement should hold for each such free
index. (If there are no free $i$ indices then this is just the
reduction to ${\cal F}^0_s$).

Functions $F(\phi ^i,\phi ^*_i)$ are now expanded as follows
\begin{equation}
F(\phi^i ,\phi ^*_i)=\sum_{m,n,p=0} \left( y_a\right)^m
\left( y_i\right)^n\left( \phi ^*_i\right) ^p F_{m,n,p}(\phi )\
,\label{Fexpmnp} \end{equation}
where $F_{m,n,p} \in  {\cal F}^1_s$, and
it has $m$ indices of type $a$
and $n+p$ indices of type $i$, which are not written.
This restriction on the coefficients is possible because if the
expansion would have
a coefficient $F_{m,n,p}$ which does not belong to
${\cal F}^1_s$, then $F_{m,n,p}-F^s_{m,n,p}$, where
$F^s_{m,n,p}\in {\cal F}^1_s$,  is
either proportional to a field
equation (and can thus be absorbed in terms of higher value of $n$),
or proportional to $R^i{}_{a}$.
In the latter case, if $i$ is one of the middle
indices, the difference can be omitted in
\eqn{Fexpmnp}. If $i$ belongs to the third set of indices this
difference can be written using an extra $y_a$ factor.

The remaining indefiniteness of this expansion is that
$y_a F^a_{m,n,p}$ could be rewritten as $\phi ^*_i G^i$, for a certain
$G^i$ of the form
$G^i=\sum_{n'=0}(y_j)^{n'}G^i_{m-1,n+n',p+1}$, where the
coefficients belong to ${\cal F}^1_s$. However, as
$\dkt$ on the first expression gives zero, $y_iG^i=0$. Therefore, as
we suppose that $G^i$ is not proportional to $R^i{}_{a}$, it follows
from \eqn{allsym} that $G^i=y_j
G^{ji}$ (antisymmetric in $[ij]$). This ambiguity thus only occurs if
$R ^i{}_a F^a \approx 0$, which implies,
due to the second line of \eqn{properZ}, that $F^a\approx
Z^a{}_{a_1}\epsilon ^{a_1}$. The ambiguous expression is thus
$y_aZ^a{}_{a_1}$.

For the functions of $\phi^i ,\phi ^*_i$ and $c^*_a$, we will use an
expansion
\begin{equation}
F(\phi^i ,\phi ^*_i, c^*_a)=\sum_{m,n=0} \left( c^*_a\right) ^m \left(
y_a\right)^n F_{m,n}(\phi ,\phi ^*)\ ,\label{Fexpa} \end{equation}
where $F_{m,n}$ has $m+n$ extra indices of type $a$ which are not
explicitly written here, and compared with \eqn{Fexpmnp}, we have
performed here the sum over $n$ and $p$. The previous expansion
\eqn{Fexpmnp} was just necessary to define this one in a proper way.
As explained above, this expansion is not unique: by \eqn{RZisf} we
have \begin{equation}
\phi ^*_i R^{i}{}_{a}Z^a{}_{a_1}-2\phi ^*_i y_j f^{ji}{}_{a_1}=0\ .
\label{phistRZ}\end{equation}
The first term appears at level $n=1$ when we expand as above. The
second term is of order zero. Indeed, if $f^{ij}$ would have terms
proportional to $R^i{}_{a}$, we can remove them by canonical
transformations (lemma~\ref{lemctrfij}). We have shown that if the
KT differential
is acyclic in one set of variables, then it is also true for the
canonically transformed variables (theorem~\ref{lemKTctr}).
The equation \eqn{phistRZ} gives an indefiniteness in the expression
for the derivative w.r.t. $y_a$.

We define an operator $d$ (unrelated to the one of the previous lemma)
\begin{equation}
dF=c^*_a \frac{\dl}{\partial y_a}F + c^*_a
Z^a{}_{a_1} \epsilon ^{a_1}\ .
\end{equation}
As explained, the derivative in the first term is not unique, and
the form of this indefiniteness can be seen by applying $d$ to
\eqn{phistRZ}.

On the other hand, the KT differential for these functions is
\begin{equation}
\dkt = y_i\frac{\dl}{\partial \phi ^*_i}
+ y_a\frac{\dl}{\partial c ^*_a} \ .
\end{equation}
We will consider the terms of \eqn{Fexpa} in decreasing order of
$n+m$. So we define
\begin{equation}
F_{(p)}=\sum_{m=0}^{p} (c^*_a)^m (y_a)^{p-m}F_{m,p-m}
\end{equation}
which leads to
\begin{equation}
(\dkt d + d \dkt)F_{(p)} =pF_{(p)} + c^*_a Z^a{}_{a_1}\epsilon
^{a_1}+G_{(p-1)}\ , \label{ddel}\end{equation}
where $\epsilon ^{a_1}$ and $G_{(p-1)}$ are undetermined. The latter,
of order at most $p-1$ in the expansion variables $y_a$ and $c^*_a$,
is produced by
\begin{equation}
\dkt \left(c^*_a Z^a{}_{a_1}\right)= 2\phi ^*_i y_j f^{ji}{}_{a_1}\ .
\end{equation}
For future use, it
is more appropriate to have the indefiniteness
proportional to a KT--invariant term. This can be done by changing
the terms of order $p-1$ with a term $(-)^i\phi ^*_i\phi ^*_j
f^{ji}{}_{a_1}\epsilon^{a_1}$. We then have
\begin{eqnarray}
(\dkt d + d \dkt)F_{(p)} &=&pF_{(p)} + y_{a_1} \epsilon ^{a_1}+ G_{(p-1)}
\nonumber\\
y_{a_1}&\equiv &c^*_a Z^a{}_{a_1}+(-)^i\phi ^*_i \phi ^*_j f^{ji}{}_{a_1}
\label{ddel2}\end{eqnarray}

Consider now a function $F\in {\cal F}^*_2$ with $\dkt F=0$.
Suppose that in the expansion of the previous lines,  the highest
level is $(p)=(N)$.
Then, as the operator $\dkt$ and $d$ do not raise the level $(p)$,
$\dkt F_{(N)}$ is of lower level, and $d\dkt F_{(N)}$ as well, except
for the indeterminacy mentioned before, which is taken into account in
\eqn{ddel2}.
Therefore this formula implies
\begin{equation}
F = \dkt H + y_{a_1}\epsilon^{a_1} +G_{(N-1)}\ , \label{Fisdkt2}
\end{equation}
where the undetermined quantities $\epsilon^{a_1}$ and $G_{(N-1)}$ are
not the same as those above. The latter has terms of level
$p\leq N-1$. Applying $\dkt$ to this expression shows that $ \dkt
G_{(N-1)}$ is proportional to $y_{a_1}$. Then
$d\dkt G_{(N-1)}$ gives zero modulo $c^*_a Z^a{}_{a_1}$. So
we may repeat the previous step, obtaining finally \eqn{Fisdkt2} with
$G_{(N-1)}$ replaced by $G_{(0)}$ of order 0, which is a function of
$\phi $ and $\phi ^*$ only, having no terms proportional to $\phi ^*_i
R^i{}_{a}$.
The KT operation on these functions can not produce $c^*_a$, and as
$\dkt G_{(0)}$ could only be proportional to $y_{a_{1}}$, we have
$\dkt G_{(0)}=0$, and we can apply the previous lemma, to obtain
\begin{equation}
F = \dkt H + y_{a_1}\epsilon^{a_1} +f(\phi )\ , \label{Fisdkt20}
\end{equation}
where $f(\phi )\in {\cal F}^0_{s}$. \QED

\begin{lemma} :
The KT differential is also acyclic on ${\cal F}^*_{3}$ (antifields
up to $c^*_{a_1}$), modulo terms
proportional to
\begin{equation}
\dkt c^*_{a_2}=c^*_{a_1}Z^{a_1}{}_{a_2}+M_{a_2}(\phi ^i,\phi
^*_i,c^*_a)\equiv y_{a_2}\ . \end{equation}
\label{lem:ac3}\end{lemma}
The proof of this lemma is very similar to that of the previous lemma.
However, some steps are a bit different here because some indices $i$
of the previous lemma remain (the zero modes are weak relations) while
some are replaced by indices $a$. We therefore present again the first
part of the proof (where differences occur), and at the end we can
refer to the previous proof.

We will expand functions
$F(\phi^i ,\phi ^*_i, c^*_a,c^*_{a_1})$ in powers of $c^*_{a_1}$ and
$y_{a_1}$.
The first expansion is obvious. The expansion in $y_{a_1}$ needs more
care. We have to choose a representant
between the antifields $c ^*_a$ to define when we have terms
proportional $y_{a_1}$.

The relation $y_a(F^a-G^a)\approx 0$ defines again an equivalence relation
between functions $F^a(\phi,\phi ^*_i)$ with a free index of type $a$.
The properness condition is the second line of \eqn{properZ}:
\begin{equation}
\mbox{If }F^a\sim G^a\mbox{ then }F^a-G^a=Z^a{}_{a_1}\epsilon
^{a_1}+y_iK^{ia}\label{FsG2}\ , \end{equation}
where $\epsilon $ and $K$ are undetermined.
The above procedure
defines for each equivalence class one representative function.
We denote the set of these representative functions as ${\cal
F}^2_{s}$. We thus obtain
\begin{equation}
y_a(F^a-G^a)\approx 0\ \mbox{ and } F^a,G^a \in {\cal F}^2_{s}\
\Longrightarrow\ F^a=G^a\ ,
\end{equation}
which again can be generalized to functions with more free indices of
type $a$.

Functions $F(\phi ^i,\phi ^*_i,c^*_a)$ are now first expanded as
follows \begin{equation}
F(\phi^i ,\phi ^*_i,c^*_a)=\sum_{m,n,p=0} \left( y_{a_1}\right)^m
\left( y_i\right)^n\left( c ^*_a\right) ^p F_{m,n,p}(\phi,\phi ^* )\
,\label{Fexpmnp2} \end{equation}
where $F_{m,n,p} \in  {\cal F}^2_s$, and
it has $m$ indices of type $a_1$, $n$ indices of type $i$,
and $p$ indices of type $a$, which are not written.
This restriction on the coefficients is possible because if the
expansion would have
a coefficient $F_{m,n,p}$ which does not belong to $
{\cal F}^2_s$, then it is in the equivalence class of some function
$F^s_{m,n,p}\in {\cal F}^2_s$, and by \eqn{FsG2},
$F_{m,n,p}-F^s_{m,n,p}$  is
either proportional to a field
equation (and can thus be absorbed in terms of higher value of $n$),
or proportional to $Z^a{}_{a_1}$.
In the latter case,  this
difference can be written using an extra $y_{a_1}$ factor.

The remaining indefiniteness of this expansion is that
$y_{a_1} F^{a_1}_{m,n,p}$ could be rewritten as $c ^*_a G^a$, for a
certain $G^a$ of the form
$G^a=\sum_{n'=0}(y_j)^{n'}G^i_{m-1,n+n',p+1}$, where the
coefficients belong to ${\cal F}^2_s$. However, taking the derivative
w.r.t. $c^*_a$, we obtain $Z^a{}_{a_1}F^{a_1}_{m,n,p}=G^a$. In the
equivalence relation which we defined, the l.h.s. is equivalent to
zero, and the r.h.s. is equivalent to the $n'=0$ term in the expansion
of $G^a$. Therefore, $G^a$ is  proportional to $y_i$.
This ambiguity thus only occurs if
$Z ^a{}_{a_1} F^{a_1} \approx 0$, which implies,
due to the properness conditions, that $F^{a_1}\approx
Z^{a_1}{}_{a_2}\epsilon ^{a_2}$. The ambiguous expression is thus
$y_{a_1}Z^{a_1}{}_{a_2}$.

We again take the sum over $n$ and $p$ and write
\begin{equation}
F(\phi^i ,\phi ^*_i, c^*_a,c^*_{a_1})=\sum_{m,n=0} \left(
c^*_{a_1}\right) ^m \left(
y_{a_1}\right)^n F_{m,n}(\phi ,\phi ^*,c^*)\ .\label{Fexpa2}
\end{equation}
The non--uniqueness of this expansion is due to
\begin{equation}
Z^{a}{}_{a_1}Z^{a_1}{}_{a_2}- y_i f^{ia}{}_{a_2}=0\ ,
\end{equation}
for some functions $f^{ia}{}_{a_2}(\phi )$. Therefore
\begin{equation}
y_{a_1}Z^{a_1}{}_{a_2} -(-)^i\phi ^*_i\phi ^*_j f^{ji}_{a_1}Z^{a_1}{}_{a_2}
-c ^*_a y_i f^{ia}{}_{a_2}=0\ .
\label{yZ2}\end{equation}
The first term appears at level $n=1$ when we expand as above. The
other
terms are of level zero. Indeed, if $f^{ia}$ would have terms
proportional to $Z^a{}_{a_1}$, we can remove them by a similar
canonical transformations as in lemma~\ref{lemctrfij}.

We can then define a new operator $d$, and
from here on there is no difference between this proof and the previous
one, replacing indices $a$ there with $a_1$, ... .\QED

The proofs for higher levels are similar to the last one.

\section{Proofs of the antibracket cohomology.}
\label{app:ABcoho}
\setcounter{equation}{0}
\setcounter{lemma}{0}
\setcounter{theorem}{0}
We first make a remark.
It is clear from \eqn{cSFexp} that if ${\cal S}F=0$, then $\dkt D^n
F=0$. But later
we will need the latter equation when we only know $\left( {\cal S}F\right)
^i=0$ for $i<n$. This is proven as follows. Define
$F_n\equiv \sum_{k=0}^{n}F^k$, so omitting from $F$ the terms $F^i$
with $i>n$. Now the nilpotency property ${\cal S}\left( {\cal
S}F_n\right)   =0$, gives at antifield level $n-1$
\begin{equation}
(-)^{F+1}\dkt \left( {\cal S}F_n\right) ^n + D^{n-1}{\cal S}F_n=0\ .
\label{idCSFn}\end{equation}
The last term depends at most on $({\cal S}F_n)^{n-1}$. As
$({\cal S} F)^i$ for $i<n$ does not depend on $F_i$ for $i>n$ (see
\eqn{cSFexp}), we can replace here $({\cal S}F_n)^i$ by $({\cal
S}F)^i$. So the last term vanishes if $\left( {\cal S}F\right) ^i=0$ for
$i<n$. On the other hand, \eqn{cSFexp} gives
$\left( {\cal S}F_n\right) ^n = D^nF$.
So we obtain
\begin{equation}
\mbox{If }\ \left( {\cal S}F\right) ^i=0  \ \mbox{for
}i< n\ \Rightarrow \dkt D^nF =0\ . \label{DnFclosed}
\end{equation}

With these results we can obtain the theorems about the cohomology. The
first one is
\begin{theorem} :  The antibracket cohomology on local
functions of negative ghost number is empty.\label{thm:ABcohoneg}
\end{theorem}
If $F$ has ghost number $f<0$ then its lowest
antifield level is $-f\geq 1$. We start from ${\cal S}F=0$ at antifield
number $-f-1$. From \eqn{cSFexp} at $n=-f-1$, we find
$\dkt F^{-f}=0$. Therefore $F^{-f}=(-)^G\dkt
G^{-f+1}$ for some local $G^{-f+1}$. Then we also have
\begin{equation}
\left( {\cal S}G^{-f+1}\right) ^{-f}=(-)^G\dkt G^{-f+1}= F^{-f}\ ,
\end{equation}
and ${\cal S}G^{-f+1}$ has no terms of lower antifield number.
Now consider
\begin{equation}
F'\equiv F-{\cal S}G^{-f+1}\ .
\end{equation}
It satisfies ${\cal S}F'=0$ and starts at antifield number $-f+1$. So we
can repeat the previous steps, constructing a $G^{-f+2}$ such that
$F-{\cal S}G^{-f+1} -{\cal S}G^{-f+2}$
starts at antifield number $-f+2$. Continuing in this way
we obtain perturbatively a function $G$ such that any local KT--closed
function $F$ of negative antifield number is equal to ${\cal S}G$.\QED
\begin{lemma} : $D^0$ is a weak nilpotent operator, which preserves the
weak
relations: if $M\approx 0$ then $D^0M\approx 0$. Therefore we can define a
weak cohomology $H^p(D^0 )$,
on functions of fields with pureghost number $p$, by the `weak relations'
\begin{equation}
D^0  M\approx 0, \qquad M\sim M'\approx M+D^0  N \ .
\end{equation}
\label{lem:nilpD0}
\end{lemma}
The nilpotency follows from
\begin{eqnarray}
D^0 D^0 F^0 &=&\left.\left( \left.(F^0,S)\right|_{\Phi ^*=0},S\right)
\right|_{\Phi ^*=0} \nonumber\\
&=&\left.\left( (F^0,S),S\right) \right|_{\Phi ^*=0} +
\left. (F^0,S)\dr{}^A\right|_{\Phi ^*=0}\cdot \left.\dl_A S
\right|_{\Phi ^*=0} \approx 0\ .
\end{eqnarray}
Indeed the last factor is only non--zero for the classical action, and this
$A$ index is thus automatically restricted to $i$.

The second statement follows from the fact that $D^0$ acts on classical
fields
as gauge transformations, and that field equations transform to field
equations under gauge transformations. Indeed
\begin{equation}
D^0 \phi ^i = R^i{}_{a}c^a\ , \label{D0phii}
\end{equation}
and \begin{equation}
M\approx 0\ \ \Rightarrow\ \ M=M^i(\dl_iS^0)\ \ \Rightarrow \ \
D^0M\approx M^i(\dl_iS^0\dr_j)R^j{}_{a}c^a\approx M^i\dl_i
\left( (S^0\dr_j) R^j{}_{a}c^a\right)  =0\ .\end{equation}\QED
We will show that this $D^0$ cohomology is equivalent to the antibracket
cohomology. We have first the following lemma
\begin{lemma} : Consider a local function $F^0$ of
antifield number 0 and of non--negative ghost number.
If it satisfies $D^0 F^0\approx 0$,
a function $F$ exists satisfying ${\cal S}F=0$ which has
$F^0$ as its part of antifield number 0.  Any 2 solutions $F$ and $F'$ of
these requirements differ by ${\cal S}G$ for a function $G$ with terms of
antifield number 2 and higher.   \label{lem:F0F1SF}
\end{lemma}
If $F^0$ satisfies $D^0  F^0\approx 0$, then \eqn{cSF00} allows a
solution for $F^1$ due to the acyclicity of the KT differential.
This implies $({\cal S}F)^0=0$ for $F=F^0+F^1+$terms with antifield
number $\geq 2$. Then \eqn{DnFclosed} implies that $\dkt D^1 F=0$. The
acyclicity property implies that there exist a function $F^2$
such that $D^1 F=(-)^{F+1}\dkt F^2 $, and thus with \eqn{cSFexp}, we have
$({\cal S}F)^1=0$.  This procedure can be iterated to find $F$ completely.

If $F^0$ is given, the vanishing of \eqn{cSFexp} for $n= 0$ (or
\eqn{cSF00}) determines $F^1$ up to $\dkt G^2$ for an arbitrary $G^2$.
With \eqn{cSFexp} we obtain
\begin{equation}
F'^1-F^1=\dkt G^2=\left( (-)^G{\cal S}G^2\right) ^1 \ .
\end{equation}
Then $F'-F-(-)^G{\cal S}G^2$ is closed under ${\cal S}$ and has its
first
term at antifield number 2. Therefore \eqn{cSFexp} on this function at
$n=1$ implies that $\left( F'-F-(-)^G{\cal S}G^2\right) ^2$ is
KT closed and the procedure can be continued to construct $G$
completely to obtain
\begin{equation}
\mbox{If }F'^0=F^0\ \mbox{ and }\ {\cal S}F={\cal S}F'=0\
\Rightarrow \ F'=F+{\cal S}G\ \mbox{ with }afn(G)\geq 2.
\end{equation} \QED
The previous lemma makes the following theorem easy.
\begin{theorem} : The antibracket cohomology at non--negative ghost number
is equal to the (weak) cohomology of $D^0 $. \label{thm:cohoabD0}
\end{theorem}
First let us make clear what we have to prove.
\begin{enumerate}
\item For any $F^0(\phi,c) $ with $D^0 F^0\approx 0$ there is a
$F(\Phi ^*,\Phi )$ such that ${\cal S}F=0$.
\item If $F'^0-F^0\approx D^0 G^0$ then $F'-F={\cal S}G$ for a local
function $G$, where $F$ and
$F'$ are the functions constructed in the first step from $F^0$ and
$F'^0$.
\item If ${\cal S}F=0$, then one can find an $F^0$ with $D^0F^0\approx
0$, and this relation is the inverse of the one in point 1.
\item If ${\cal S}F={\cal S}F'=0$ and $F-F'={\cal S}G$, then
$F^0-F'^0\approx D^0 G^0$ for $F^0$ and $F'^0$ defined in the previous
steps.
\end{enumerate}
Lemma~\ref{lem:F0F1SF} made already the construction of $F$ mentioned
in point 1, and the inverse (point 3) of this is just its restriction
to antifield number 0. The equation $D^0 F^0\approx 0$ follows then from
\eqn{cSF00} using that $\dkt$ on terms of antifield number 1 produces field
equations. For point 2, we have that $F'^0-F^0-D^0 G^0$ is weakly
vanishing and thus equal to $\dkt G^1$ for some local $G^1$. Therefore
\begin{equation}
F'^0=F^0+D^0 G^0+(-)^G\dkt G^1=F^0+\left( {\cal S}(G^0+G^1)\right) ^0\
. \end{equation}
As $F'$ and $F+{\cal S}(G^0+G^1)$ are both closed under ${\cal S}$ and
equal at antifield number 0, it follows from the second part of the
lemma that they differ by ${\cal S}$ on some function starting at
antifield number 2, which proves point 2. Finally for point 4, the
assumption at antifield number 0 reads
\begin{equation}
F^0-F'^0=(-)^G\dkt G^1 +D^0 G^0\approx D^0 G^0\ .
\end{equation}
\QED

\end{document}